\documentclass[a4paper,fleqn,usenatbib]{mnras} 

\usepackage{newtxtext,newtxmath}

\usepackage[T1]{fontenc}
\usepackage{ae,aecompl}

\usepackage{graphicx}	
\usepackage{amsmath}	
\usepackage{amssymb}	

\newcommand*{\Vect}[1]{\ensuremath{\boldsymbol{#1}}}
\newcommand*{\Grad}{\Vect{\nabla}}
\newcommand*{\Div}{\Vect{\nabla}\cdot}
\newcommand*{\Curl}{\Vect{\nabla}\!\times\!}
\newcommand*{\Sh}{\Delta^*}

\title[Magnetic field evolution in neutron star cores]{Magnetic field evolution and equilibrium configurations in neutron star cores: the effect of ambipolar diffusion}

\author[Castillo et al.]{
F. Castillo,$^{1}$\thanks{E-mail: fcastillo21@um.uchile.cl}
A. Reisenegger,$^{2}$
and J. A. Valdivia$^{1,3}$
\\
$^{1}$Departamento de F\'{\i}sica, Facultad de Ciencias,
Universidad de Chile, Santiago, Chile.\\
$^{2}$Instituto de Astrof\'isica, Pontificia Universidad Cat\'olica de Chile, Santiago, Chile.\\
$^{3}$Centro para el Desarrollo de la Nanociencia y
Nanotecnolog\'\i a, CEDENNA, Santiago, Chile.
}

\date{Accepted XXX. Received YYY; in original form ZZZ}

\pubyear{2016}

\begin{document}
\label{firstpage}
\pagerange{\pageref{firstpage}--\pageref{lastpage}}
\maketitle

\begin{abstract}
As another step towards understanding the long-term evolution of the magnetic field in neutron stars, we provide the first simulations of ambipolar diffusion in a spherical star. Restricting ourselves to axial symmetry, we consider a charged-particle fluid of protons and electrons carrying the magnetic flux through a motionless, uniform background of neutrons that exerts a collisional drag force on the former. We also ignore the possible impact of beta decays, proton superconductivity, and neutron superfluidity. All initial magnetic field configurations considered are found to evolve on the analytically expected time-scales towards ``barotropic equilibria'' satisfying the ``Grad-Shafranov equation'', in which the magnetic force is balanced by the degeneracy pressure gradient, so ambipolar diffusion is choked. These equilibria are so-called ``twisted torus'' configurations, which include poloidal and toroidal components, the latter restricted to the toroidal volumes in which the poloidal field lines close inside the star. In axial symmetry, they appear to be stable, although they are likely to undergo non-axially symmetric instabilities.
\end{abstract}

\begin{keywords}
stars: neutron -- stars: magnetic field -- MHD -- methods: numerical
\end{keywords}



\section{Introduction}

The structure and evolution of the magnetic fields in neutron stars
(NSs) is an interesting and current topic of research. For instance, there seems to be a pronounced
difference in the magnetic field strength among the different classes
of neutron stars, which appears to be correlated with
their age. Young neutron stars like radio pulsars and high-mass X-ray
binaries have a surface magnetic field of the order of $10^{12}$G,
while older millisecond pulsars have lower fields of the order of
$10^{9}$G, which can be explained in terms of accretion
\citep{Bhattacharya1995}, although recent works have explored the
possibility that this field decay may be explained in terms of
intrinsic mechanisms of the stars (Cruces, Reisenegger, \& Tauris, in
preparation).  Objects with ultrastrong $10^{14-15}$G magnetic fields
have also been observed as anomalous X-ray pulsars and soft gamma
repeaters, collectively known as magnetars, whose strong activity can
only be explained in terms of a decaying magnetic field, since their
radiated energy is much larger than their rotational energy loss.
Since these objects appear to be isolated, their field decay must be
attributed to processes intrinsic to the neutron stars
\citep{Thompson1995, Thompson1996}.  For a review of the different
classes of neutron stars and their properties see \citet{Kaspi2010}.
Understanding the mechanisms that drive the evolution of the magnetic
field in neutron stars may be crucial to understand the relation
between these kinds of objects.

In the solid crust of NSs, ions have very restricted mobility and currents are carried by electrons.  As discussed by \citet{Goldreich1992a}, the
long-term mechanisms that promote the evolution of the field in this
region are Ohmic diffusion and Hall drift. Ohmic diffusion is a
dissipative effect caused by electric resistivity, which depends on a very uncertain ``impurity
parameter'' \citep{Cumming2004}.  Hall drift, on the other hand, is an
advection of the magnetic field by the electrical current. Although not a dissipative effect, it might cause a
rearrangement of the magnetic field on astrophysically relevant
time-scales.  Numerous studies have focused on understanding the
effects of Hall drift, however, due to the non-linearity
of the equations involved, these studies have mainly taken a numerical
approach \citep{Hollerbach2002, Pons2007, Pons2009, Vigano2012,
  Gourgouliatos2013, Vigano2013a}, finding that the magnetic field evolves towards stable ``attractor'' configurations
\citep{Gourgouliatos2014a, Marchant2014}.

The core of a NS is a fluid mixture of neutrons, protons, and electrons, joined by other species at increasing densities. Its density and conductivity are higher than those of the crust, making Hall drift and Ohmic diffusion substantially slower. On the other hand, since all particles can move, the magnetic flux can in principle be transported and re-arranged by convective motions. However, as the density profiles of neutrons and charged particles are different, their joint radial motions are opposed by strong buoyancy forces \citep{Pethick1992,Goldreich1992a}, unless the composition can be adjusted in ``real time'' by converting one species to another by weak interaction processes (so-called ``Urca reactions''). These are strongly temperature-dependent, making this feasible only during a very short time after the neutron star birth. After that, and as long as the different species are strongly coupled by collisions, they can be treated as a single, stably stratified, non-barotropic fluid. However, as the temperature decreases, the collisional coupling is reduced, making relative motions possible, and thus allowing the charged particles to carry the magnetic flux without dragging the neutrons along with them, in a process called ``ambipolar diffusion''  \citep{Pethick1992,Goldreich1992a}. Hence, on long time-scales, charged particles
(coupled to the magnetic field) and neutrons act as independent barotropic fluids (if muons or charged particles other
than $p$ and $e$ can be ignored).

Ambipolar diffusion has been invoked to explain the activity of magnetars
due to its strong dependence on the magnetic field intensity
\citep{Thompson1995, Thompson1996}. It may also be important to
explain the low magnetic fields of millisecond pulsars, as the collisional coupling between charged particles and neutrons decreases at low temperatures, so the timescale of ambipolar diffusion in millisecond pulsars may be short enough to predict substantial magnetic field decay in old neutron stars prior to their stage of accretion (Cruces, Reisenegger, \& Tauris, in preparation).
Hence, it is relevant to further study the impact of this effect on the long-term evolution of the magnetic field, as well as to check if it will lead to a complete decay of the magnetic field or to a stable magnetic equilibrium.

In a purely fluid star, stable magnetic equilibria appear to require
stable stratification (i.e., stability against convective motion by a
gradient of entropy or composition) and a combination of poloidal
and toroidal field components balancing each other
\citep{Prendergast1956,Braithwaite2004,Braithwaite2006a,Reisenegger2001,Reisenegger2009,Akgun2013,Mitchell2015}.
This is in agreement with MHD simulations performed by
\citet{Braithwaite2004,Braithwaite2006a}, where matter within the
star is non-barotropic (i.e. pressure depends
on a second quantity, such as chemical composition or specific
entropy, in addition to density). These simulations showed that random initial fields can evolve naturally, within a few Alfv\'en times, into toroidal-poloidal, nearly axisymmetric ``twisted torus'' configurations, in which the toroidal magnetic field is confined to torus shaped regions inside the star (\citealt{Braithwaite2004}; but see also \citealt{Braithwaite2008}).

If the previous conditions are not met, the magnetic
field decays due to non-axisymmetric instabilities \citep[e.g.,][]{Mitchell2015}. Thus, if we can
ignore the NS crust (assuming it has a low conductivity), the magnetic
field in the core could be stable on time-scales much shorter than
ambipolar diffusion (as long as the core can be treated as a single,
non-barotropic fluid), but evolve as ambipolar diffusion decouples
the two barotropic components, eventually decaying through non-axisymmetric instabilities.
(The latter may not be the case if the charged component is stabilized by a muon density gradient.)

The model introduced by \citet{Goldreich1992a} considered the core of
NSs as a plasma composed of neutrons ($n$), protons ($p$), and
electrons ($e$).  These species suffer binary collisions, convert into
each other by weak interactions, and are affected by the macroscopic
magnetic field.  In their analysis, for simplicity, neutrons were
assumed to form a motionless background. This restriction was later
relaxed by \citet{Reisenegger2007a, Hoyos2010} who addressed the
problem numerically, solving the relevant equations in one dimension.

Understanding the full impact of the ambipolar diffusion on the
evolution of the magnetic field requires numerically evolving the
relevant effects in time. In this work we go back to the assumption of a motionless, uniform background of neutrons, but extend the previous work to a more realistic geometry, namely a spherical star with an axially symmetric magnetic field configuration. The latter restriction is quite constraining, as it does
not allow for some of the main magneto-hydrodynamic instabilities
identified for axially symmetric magnetic fields in stably stratified
stars, all of which break the axial symmetry and are suppressed only
in stably stratified fluids with some specific magnetic field configurations \citep{Tayler1973, Markey1973,Wright1973,Braithwaite2009, Mitchell2015}.
Similar instabilities are likely to arise in the present context, and therefore not all field configurations appearing to be stable in axial symmetry will remain so if general 3-dimensional motions are allowed.
Addressing 3-dimensional motions, the motions of neutrons, as well as the high temperature regime (in which weak interactions between species become relevant) will be left for a future manuscript.

In the present work we have also neglected the effects of superfluidity and superconductivity. Including these requires an understanding of non-trivial micro-physics and might change time-scales and equilibrium configurations. Several authors have addressed this issue in the last years (e.g., \citealt{Glampedakis2011,Graber2015,Elfritz2016,Passamonti2017}), however they still have not reached consensus (see the criticisms by \citealt{Gusakov2016, Dommes2017}, and the response of \citealt{Passamonti2017b}). The impact of superfluidity and superconductivity seems to be well understood only in particularly simple situations, such as the one studied by \citet{Kantor2017}. Certainly, the recent work of \citet{Kantor2017, Passamonti2017b}, and particularly \citet{Gusakov2017}, present a step forward in this topic, however they still lack some key ingredients eventually needed to evolve the magnetic field consistently in the presence of superfluidity/superconductivity. Among these are the dynamics and interactions of quantized neutron vortices and magnetic flux tubes, and the possible presence of alternating layers in which superfluidity or superconductivity is present and others in which it is not.

This paper is organized as follows. In Sec.~\ref{sec:model} we discuss
the physical model and construct the equations to be solved in axial
symmetry. Here we also discuss the equation satisfied by eventual
equilibrium configurations as well as the relevant time-scales.  In
Sec.~\ref{sec:results_timescales} we check if the simulations are in
agreement with the time-scales analytically expected.  In
Sec.~\ref{sec:results_equilibriumconfigurations} we study the
equilibrium configurations found, and check if they are solutions of
the GS equation.  The stability of solutions of this equation is
studied in Sec.~\ref{sec:results_stabilityofequilibria}. Finally, in
Sec.~\ref{sec:conclusions}, our results are summarized and conclusions are
outlined.

\section{Physical model}
\label{sec:model}

Based on the model described by \citet{Goldreich1992a} and
\citet{Reisenegger2007a}, we model the interior of an isolated neutron
star as a plasma composed of neutrons, protons, and electrons,
considering the simplified scenario in which the neutrons form a
motionless background. The charged species are coupled by collisions
and electromagnetic forces, and their equations of motion are written
as
\begin{gather}
\begin{split}
	n_c\frac{\mu_p}{c^2}\frac{d\Vect{v_p}}{dt}=&+ n_c e\left(\Vect E +\frac{\Vect{v_p}}{c}\times\Vect B \right) -n_c\Grad\mu_p - \frac{n_c\mu_p}{c^2}\Grad\Psi\\
	& - \gamma_{pe}n_c^2 \left( \Vect{v_p} - \Vect{v_e}\right) - \gamma_{pn}n_c n_n\Vect{v_p} \,,\label{eq:protones}
\end{split}\\
\begin{split}
	n_c\frac{\mu_e}{c^2}\frac{d\Vect{v_e}}{dt}=&- n_c e\left(\Vect E +\frac{\Vect{v_e}}{c}\times\Vect B \right) -n_c\Grad\mu_e - \frac{n_c\mu_e}{c^2}\Grad\Psi\\
	&  - \gamma_{ep}n_c^2 \left( \Vect{v_e} - \Vect{v_p}\right) - \gamma_{en}n_c n_n\Vect{v_e} \,,\label{eq:electrones}
\end{split}
\end{gather}
where $n_i$ and $\mu_i$ ($i=p,e$) are the number density and chemical
potential of species $i$, respectively; $\mu_i/c^2$ is the effective
mass of each species, which could include corrections due to
interactions and relativistic effects \citep{Akmal1998}; $\pm e$ are
the charges of protons and electrons, respectively, and $\Vect{v_i}$ is
the velocity of the $i^{th}$ species. We assume charge neutrality, so
that at all times $n_p=n_e\equiv n_c$.

The forces acting on each particle are, from left to right, the
Lorentz force (where $\Vect E$ and $\Vect B$ are the electric and
magnetic fields), the degeneracy-pressure gradient of species $i$, the
gravitational force acting on each particle (where $\Psi$ is the
gravitational potential), and the frictional
drag forces due to collisions between particles of different species,
parametrized by the parameter $\gamma_{ij}$,
which is inversely proportional to the mean collision time between
species of type $i$ and $j$ (where $n$ denotes the neutrons).

We assume that, under a given perturbation, the
star very quickly reaches a magneto-hydrostatic quasi-equilibrium
state in which all the forces on a fluid element are close to
balancing each other. The time-scale to reach this state is a few
Alfv\'en times, and, since the process under
consideration involves small velocities that change over scales much
longer than the Alfv\'en time, we rewrote the equations in a slow-motion
 approximation that neglects the inertial terms on the left
side of the equations of motion \citep{Reisenegger2007a}. Hence, by solving these
equations for the electric field $\Vect{E}$ and using Faraday's law, it
is possible to get an evolution equation for the magnetic field given by
\begin{gather}
\begin{split}
\frac{\partial\Vect B}{\partial t} =& \Curl\left[\left( \Vect{v_A}+\Vect{v_H}\right)\times\Vect B -\frac{c}{\sigma}\Vect{J}\right]\\ &-\frac{c}{e}\Grad\left(\gamma_{en}+\gamma_{pn}\right)^{-1}\times\Grad\left(\mu_p\gamma_{en}-\mu_e\gamma_{pn}\right)\\ &-\frac{1}{ec}\Grad\left(\frac{\mu_p\gamma_{en}-\mu_e\gamma_{pn}}{\gamma_{en}+\gamma_{pn}}\right)\times\Grad\Psi \,.
\end{split}\label{eq:induccion}
\end{gather}
Here, we see how the magnetic field lines will drift with the ``ambipolar diffusion velocity'', namely
\begin{equation}
\Vect{v_A} = \frac{\gamma_{pn}\Vect{v_p}+\gamma_{en}\Vect{v_e}}{\gamma_{en}+\gamma_{pn}} \,, \label{eq:def_va}
\end{equation}
which represents the joint motion of the two charged particle species; and the Hall drift velocity $\Vect{v_H}=-(\gamma_{pn}-\gamma_{en})(\Vect{v_p}-\Vect{v_e})/(\gamma_{en}
+ \gamma_{pn})$, which is proportional to the electric current density $\Vect{J}=n_ce(\Vect{v_p}-\Vect{v_e})=c\Curl\Vect{B}/4\pi$. The
next term represents Ohmic decay, where
\begin{equation}
	\sigma = e^2\left(\gamma_{pe}+\frac{\gamma_{en}\gamma_{pn}}{\gamma_{en}+\gamma_{en}}\frac{n_n}{n_c} \right)^{-1} \,,
\end{equation}
is the electric conductivity. The last two terms represent battery effects.
In this work we are interested in studying the effects and dynamics of ambipolar diffusion in the core of NSs.
As shown by \citet{Goldreich1992a}, the time-scale of ambipolar diffusion is
\begin{equation}
t_\text{ambip} \sim 3\times 10^{7} \frac{L_6^2 T_6^2}{B^2_{12}} \,\text{yr}\,, \label{eq:t_ambip}
\end{equation} 
where $B_{12}\equiv B/(10^{12}\text{G})$, $L_6\equiv L/(10^{6}\text{cm})$, and $T_6\equiv T/(10^{6}\text{K})$, which in old NSs can be orders of magnitude faster
than the effects of Hall drift and Ohmic decay, namely,
\begin{gather}
t_{\text{Hall}} \sim 5\times 10^{10} \frac{L_6^2}{B_{12}}\left(\frac{\rho}{\rho_{\text{nuc}}}\right) \,\text{yr}\,, \label{eq:thall}\\
t_{\text{Ohm}} \sim 2\times 10^{17} \frac{L_6^2}{T_6^2}\left(\frac{\rho}{\rho_{\text{nuc}}}\right)^3 \,\text{yr} \,,\label{eq:tohm}
\end{gather}
where $\rho$ is the mass density of the star and, $\rho_{\text{nuc}}\equiv2.8\times 10^{14}\,\text{g}\,\text{cm}^{-3}$, so that we neglect the latter effects in the induction equation. We also neglect the battery terms, in agreement with \citet{Goldreich1992a} (see discussion in Sec.~\ref{sec:model_neglect_of_battery_terms}).

From equations \eqref{eq:protones}, \eqref{eq:electrones}, and \eqref{eq:def_va} we obtain
\begin{equation}
	\Vect{v_A} = \frac{1}{\gamma_{cn}n_c n_n}\left[\frac{\Vect{J}\times\Vect{B}}{c} - n_c\left(\Grad\mu_c+\mu_c\frac{\Grad\Psi}{c^2}\right)\right]\label{eq:ambipolar}\,,
\end{equation}
where $\mu_c=\mu_e+\mu_p$, and $\gamma_{cn}=\gamma_{en}+\gamma_{pn}$, showing that ambipolar diffusion is driven by the magnetic, pressure, and gravitational forces acting on the charged particles, and opposed by the collisional drag of the neutrons. To evolve the particle density, we use the continuity equation of the charged particles, namely,
\begin{equation}
	\frac{\partial n_c}{\partial t} + \Div\left(n_c\Vect{v_A}\right) = 0 \label{eq:continuidad}\,.
\end{equation}

Since the ratio between the magnetic pressure and degeneracy pressure
$P$ in the interior of NSs is $B^2/8\pi
P \lesssim 10^{-6}$, the perturbations to the particle density
profile and hydrostatic equilibrium induced by the magnetic field are
expected to be similarly small. We will thus split the particle
density, and hence the chemical potential, in two: a time-independent background density $n_c$ and chemical potential $\mu_c$ set by a
hydrostatic equilibrium state in the absence of the magnetic field, namely,
\begin{equation}
	\Grad\mu_c+\frac{\mu_c}{c^2}\Grad\Psi=0, \label{eq:eqhidrostatico}
\end{equation}
and much smaller time-dependent perturbations $\delta n_c$ and $\delta \mu_c$, respectively, induced by the evolving
magnetic field. Since the latter are small, we can linearize $\delta\mu_c = K_{cc}\delta n_c$, where $K_{cc}=d\mu_e/d
n_c + d\mu_p/d n_c$.

By dropping higher order terms in the above
expressions we can write the full set of equations we plan to solve as

\begin{gather}
	\frac{\partial\Vect B}{\partial t} = \Curl\left(\Vect{v_A}\times\Vect B\right)  \,, \label{eq:induccion0}\\
	n_c\Vect{v_A} = \frac{1}{\gamma_{cn} n_n}\left[\frac{\Vect{J}\times\Vect{B}}{c}- n_c\mu_c\Grad\left(\frac{\delta\mu_c}{\mu_c}\right)\right] \,,\label{eq:va0}\\
	\frac{\partial\delta n_c}{\partial t} + \Div\left(n_c\Vect{v_A}\right) = 0 \label{eq:continuidad0}\,,\\
	\delta\mu_c = K_{cc}\delta n_c \label{eq:dmuc0}\,,
\end{gather}
where the degeneracy pressure gradient and the gravitational force have been combined in one term using equation~\eqref{eq:eqhidrostatico}.

\subsection{Time scales for ambipolar diffusion}
\label{sec:model_timescales}

From equations \eqref{eq:induccion0} and \eqref{eq:continuidad0}, we can see that the time required for substantial changes in the magnetic field configuration is $t_B\sim R/v_A$ (where $R$ is the core radius), whereas important changes in the density perturbations can occur much faster, in $t_c\sim t_B\delta n_c/n_c$. Thus, on short time-scales $\sim t_c$, the magnetic field configuration can be regarded as fixed, while (from equations \ref{eq:va0}--\ref{eq:dmuc0}) the density perturbation satisfies a forced diffusion-type equation,
\begin{equation}
{\partial\delta n_c\over \partial t}=\nabla\cdot\left\{{1\over\gamma_{cn}n_n}\left[n_c\mu_c\nabla\left(K_{cc}\delta n_c\over\mu_c\right)-{\Vect J\times\Vect B\over c}\right]\right\} \,. \label{eq:16}
\end{equation}
Thus, the evolution time for the density perturbations can be estimated more precisely as 
\begin{equation}
t_{c} \sim \frac{\gamma_{cn} n_n R^2}{K_{cc} n_{c}} \sim 10^{-3}T_6^2\,\text{yr}\,,\label{eq:tc}
\end{equation}
on which the charged particles will diffuse through the neutrons towards a quasi-stationary state that makes the right-hand side of equation \eqref{eq:16} vanish, and in which
\begin{equation}
\delta n_c \sim \frac{B^2}{4\pi n_c K_{cc}} \,. \label{eq:dnc}
\end{equation}
As the magnetic field evolves on the longer time scale $t_B$, the density perturbations will easily keep up, always maintaining this condition to good approximation. In other words, on the right-hand side of equation \eqref{eq:va0}, the chemical potential perturbation will always adjust so as to keep $\nabla\cdot(n_c\Vect v_A)\approx 0$. Thus, except for very special magnetic field configurations, the two terms on the right-hand side will be of the same order of magnitude, but possibly have very different curls, allowing for a solenoidal flow with characteristic speed
\begin{equation}
v_A^\text{s} \sim \frac{\left|(\Curl\Vect{B})\times\Vect{B}\right|}{4\pi\gamma_{cn}n_c n_n} 
\sim \frac{B^2}{4\pi\gamma_{cn}n_c n_n R}\,,
\end{equation}
so the time scale for magnetic field evolution becomes 
\begin{equation}
t_B \sim \frac{4\pi\gamma_{cn}n_c n_n R^2}{B^2} \sim 5\times10^{7} T_6^2 B_{12}^{-2}\,\text{yr} \,. \label{eq:tb}
\end{equation}
Of course, over this time scale, the magnetic field could in principle evolve towards an equilibrium state in which the two terms on the right-hand side of equation \eqref{eq:va0} exactly cancel, making the ambipolar velocity vanish.

\subsection{Neglect of battery terms}
\label{sec:model_neglect_of_battery_terms}

The magnitude of the battery terms in equation \eqref{eq:induccion} can be estimated from the equation as $c\delta\mu_e/eR^2$, which gives us a time-scale $t_\text{batt}$ for the battery effects
\begin{equation}
\frac{B}{t_\text{batt}}\sim \frac{c\delta\mu_e}{eR^2} \sim \frac{cK_{ec}\delta n_c}{eR^2}\,,
\end{equation}
where $K_{ec}=d\mu_e/dn_c\sim K_{cc}$. The magnitude $\delta n_c$ is given from equation \eqref{eq:dnc}, and therefore
\begin{equation}
t_\text{batt} \sim \frac{B eR^2}{cK_{cc}\delta n_c} \sim \frac{4\pi e n_c R^2}{c B} \sim 2\times 10^{11} \frac{L_6^2}{B_{12}}\,.
\end{equation}
It is interesting to note that this time-scale is of the same order of magnitude as the one
obtained from Hall drift (equation \ref{eq:thall}), namely
$t_\text{Hall}\sim R/v_\text{Hall}$, where $v_\text{Hall}\sim J/n_c
e\sim cB/4\pi e n_c R $. Therefore, since $t_B/t_\text{batt} \sim t_B/t_\text{Hall}\sim 3\times 10^{-4}
T_6^2/B_{12}$, it is safe to neglect both the Hall effect and the battery terms in the previous equations.

\subsection{Dimensionless equations for a uniform background}
\label{sec:model_dimensionless_equations}

In the rest of the manuscript we will also assume, for simplicity, that the background densities $n_c$, $n_n$, and all transport parameters are spatially uniform. 
 
We have written the equations in dimensionless form,
\begin{gather}
	\frac{\partial\Vect B}{\partial t} = \Curl\left(\Vect{v_A}\times\Vect B\right)  \,, \label{eq:induccion_a}\\
	\Vect{v_A} = b^2(\Curl\Vect{B})\times\Vect{B} - \Grad\chi \,,\label{eq:va_a}\\
	\frac{\partial\chi}{\partial t} + \Div\Vect{v_A} = 0 \label{eq:continuidad_a}\,,
\end{gather}
where distances have been normalized to the radius $R$ of the core, and $n_c$, $\gamma_{cn}$, and $K_{cc}$ have been normalized to one. Chemical potentials are in units of $\mu_0=K_{cc}n_{c}$, so that $\chi\equiv\delta n_c/n_{c}=\delta\mu_c/\mu_0$. Velocities have been normalized to
$R/t_{c}$. The magnetic field is in units of $B_0$, the root mean square of the magnetic field in the volume of the star, and
\begin{equation}
	b^2 \equiv \frac{B_0^2}{4\pi K_{cc}n_{c}^2} \label{eq:b2def}\,,
\end{equation}
which is of the order of the (very small) ratio between the magnetic and degeneracy pressure. Thus, $B_0 = n_{c}\sqrt{4\pi K_{cc}}\:b \sim 2.3 \times 10^{17}b\, \text{G}$, and the dimensionless time-scales are
\begin{gather}
t_c\sim 1 \,,\\
t_B\sim \frac{1}{b^2}\,.
\end{gather}
Therefore, in order to properly resolve both time-scales in our simulations without having to use a prohibitively small time-step we will scale the magnetic field to be much higher than $10^{12}$G, so
that both time-scales $t_B$ and $t_c$ get closer to each other.

\subsection{Axially symmetric fields}
\label{sec:model_axially_symmetric_fields}

We restrict ourselves to axial symmetry, so
\begin{equation}
\Vect{B}=\Grad\alpha\times\Grad\phi + \beta\Grad\phi \,. \label{eq:BAlfaBeta}
\end{equation}
Here the scalar potentials $\alpha(t,r,\theta)$ and $\beta(t,r,\theta)$ generate
the poloidal and toroidal magnetic field, respectively, where $t$ denotes time, $r$ is
the radial coordinate, and $\theta$ and $\phi$ are the polar and azimuthal
angles, respectively, so $\Grad\phi=\hat{\phi}/(r\sin\theta)$. An explicit form for the evolution of the magnetic potentials
can be derived from equation~\eqref{eq:induccion_a}, where we get
\begin{gather}
	\frac{\partial\alpha}{\partial t}=r\sin\theta\left(\Vect{v_A}\times\Vect{B}\right)\cdot\hat\phi\,, \label{eq:evolucionalfa}\\
	\frac{\partial\beta}{\partial t}=r^2\sin^2\theta\Div\left[\frac{\left(\Vect{v_A}\times\Vect{B} \right)\times\hat{\phi}}{r\sin\theta} \right]\,. \label{eq:evolucionbeta}
\end{gather}

Note that, when decomposing the vector fields into poloidal and toroidal components, $\Vect B=\Vect {B_{\text{Pol}}}+\Vect {B_{\text{Tor}}}$ and $\Vect {v_A}=\Vect {v_{A\text{,Pol}}}+\Vect {v_{A\text{,Tor}}}$, the evolution of $\Vect {B_{\text{Pol}}}$ (equation \ref{eq:evolucionalfa}) will only depend on $\Vect {v_{A\text{,Pol}}}\times\Vect {B_{\text{Pol}}}$ (poloidal transport of the poloidal field component), whereas the evolution of $\Vect {B_{\text{Tor}}}$ (equation \ref{eq:evolucionbeta}) depends on two terms: $\Vect {v_{A\text{,Pol}}}\times\Vect {B_{\text{Tor}}}$ (poloidal transport of the toroidal field component), but also $\Vect {v_{A\text{,Tor}}}\times\Vect {B_{\text{Pol}}}$ (winding [or unwinding] of poloidal field lines into a toroidal field by a toroidal velocity field).

For the ambipolar velocity, we obtain
\begin{equation}
	\Vect{v_A} = -\frac{b^2}{r^2\sin^2\theta}\left(\Sh\alpha\Grad\alpha+\beta\Grad\beta-\Grad\alpha\times\Grad\beta\right) - \Grad\chi \,,\label{eq:va_AB}
\end{equation}
where we have introduced the Laplacian-like ``Grad-Shafranov''
operator \citep{Grad1958,Shafranov1966}, defined as $\Sh\alpha \equiv
r^2\sin^2\theta\Div(\Grad\alpha/r^2\sin^2\theta)$. 

Since $\Vect {v_{A\text{,Tor}}}\propto \Grad\alpha\times\Grad\beta$, purely poloidal fields will remain poloidal, and purely
toroidal fields will remain toroidal in time. Although there are
strong suggestions that such single-component configurations are always unstable under non-axially-symmetric
perturbations~\citep{Tayler1973,Braithwaite2009}, we will analyze (among other things) if ambipolar diffusion can produce stable equilibrium configurations with purely poloidal or toroidal magnetic fields in axial symmetry, and if they are considerably different from the combined poloidal+toroidal magnetic field configurations that may be stable.

\citet{Cumming2004} showed that the Ohmic decay time-scale in the crust of NSs is given by $t_{\text{Ohm}}\sim 5.7\text{Myr}/Q$, where $Q$ is called the ``impurity parameter''. However, the value of $Q$ is highly uncertain; it has been estimated to be as small as $10^{-3}$ \citep{Flowers1977} and as large as $10$ \citep{Jones2001}.
In addition, for sufficiently strong magnetic fields, the decay time might be shortened by the additional action of the Hall drift.
Therefore, it is unclear if currents in the core or in the crust will decay faster. In this work we will assume the latter.
Hence, we treat the crust as a vacuum, whose magnetic field at any time is fully determined by the field in the core.

\subsection{Boundary conditions}
Since we are restricted to axial symmetry, we cannot have magnetic field lines or charged particles crossing through the axis, and therefore we must have $B_\theta=B_\phi=0$, and $v_{A,\theta}=v_{A,\phi}=0$ there. The condition on the magnetic field imposes $\partial\alpha/\partial r=\beta=0$. Hence, $\alpha$ must be constant along the axis, whose value we set to zero, in which case the magnetic flux enclosed by any circle around the axis at fixed $r$ and $\theta$ can be written as $2\pi\alpha(r,\theta)$. 

We assume that in the crust and outside of the star we have a perfect vacuum, so the field outside of the core is determined by its radial component at the crust-core interface, which must be continuous.
We are also considering that the surface currents dissipate in a time-scale that is much shorter than the ones interesting to us, so the tangential component of the magnetic field must also be continuous there.

Since we assume no currents in the crust and outside the star, the magnetic field can be written as $\Vect B=\Grad\psi$ there, where $\psi$ is a scalar potential solution of $\nabla^2\psi=0$, and $\psi$ must go to zero when $r$ goes to infinity. Hence, as shown by \citet{Marchant2011}, we can match the internal field with a multipolar expansion outside, given by
\begin{equation}
	\psi(r,\theta) =\sum_{l=1}^\infty \frac{a_l}{r^{l+1}}P_l(\cos\theta)  \,,\label{multipolos}
\end{equation}
where $P_l(\cos\theta)$ is the Legendre polynomial of index $l$.
The coefficients $a_l$ are determined by the radial component of the magnetic field at the crust-core interface (of radius $R=1$) as
\begin{equation}
	a_l=-\frac{2l+1}{2(l+1)}R^{l+2}\int_0^\pi B_r(R,\theta) P_l(\cos\theta) \sin\theta  \,d\theta  \,.\label{eq:a_l}
\end{equation}
Integrating by parts, we obtain
\begin{equation}
	a_l=\frac{2l+1}{2(l+1)}\int_0^\pi \alpha(1,\theta) P_l^1(\cos\theta)\,d\theta  \,,
\end{equation}
where $P_l^1(\cos\theta)$ are the associated Legendre
polynomials of index $(l,1)$. 

Hence, at the crust-core interface, $\alpha$ and its radial derivative must be continuous, i.e., $\alpha$ must satisfy
\begin{equation}
B_\theta(1,\theta,\phi)=\frac{1}{r\sin\theta}\left.\frac{\partial\alpha}{\partial r}\right|_{r=1}=\left.\frac{1}{r}\frac{\partial\psi}{\partial\theta}\right|_{r=1}\,.
\end{equation}
As for $\beta$, since $B_\phi=0$ we set $\beta=0$ there.

Lastly, at the crust-core interface we assume that the radial component of the ambipolar velocity is null, since the positive charges in the crust are fixed to the crystal lattice.
This is achieved by imposing a strong degeneracy pressure gradient at the surface, to balance the radial magnetic force, as suggested by equation~\eqref{eq:va_a}.

\subsection{Energy stored in the magnetic field and the particles}

The magnetic energy is given by
\begin{align}
	U_B&=\frac{1}{8\pi}\int d^3x \, |\Vect{B}|^2 \,,\label{eq:UB}\\
	&= \frac{1}{8\pi}\left(\int d^3x \, |\Vect{B_\text{Pol}}|^2 +\int d^3x \, |\Vect{B_\text{Tor}}|^2\right) \,,
\end{align}
where we split the magnetic field in its poloidal and toroidal components, since they are orthogonal. In terms of the magnetic potentials $\alpha$ and $\beta$, the two terms in the last expression read 
\begin{gather}
	U_{\text{Pol,tot}}=\frac{1}{4}\int \frac{dr\,d\theta}{\sin\theta}\left[ \left(\frac{1}{r}\frac{\partial\alpha}{\partial\theta}\right)^2 + \left(\frac{\partial\alpha}{\partial r}\right)^2 \right] \,,\\
	U_{\text{Tor}}=\frac{1}{4}\int \frac{\beta^2}{\sin\theta} dr\,d\theta \,.
\end{gather}
The poloidal magnetic energy can be split into the magnetic energy stored inside of the star $U_{\text{Pol,int}}$ and the external energy stored in the magnetic field outside of the star $U_{\text{Pol,ext}}$. Noting that outside of the star $|\Vect{B}|^2=\Div(\psi\Grad\psi)$, we can use Gauss's theorem, along with the orthogonality properties of the Legendre polynomials, to get
\begin{equation}
	U_{\text{Pol,ext}}=\sum_{l=1}^\infty \frac{l+1}{2(2l+1)}a_l^2 \,.
\end{equation}

The variation of energy stored in the particles of species $i$ (i.e, in the Fermi sea; $i=p,e$) $\delta U_{\text{c,i}}$, induced by the magnetic field, can be evaluated integrating the energy of the particles with energies between the background Fermi energy and the perturbed Fermi energy of the species in the volume $V$ of the star, namely,
\begin{equation}
\delta U_{\text{c,i}}=\int_V d^3x\int_{\mu_i}^{\mu_i+\delta\mu_i(\Vect{x})}D_i(\epsilon,\Vect{x})\epsilon\, d\epsilon \,.\\
\end{equation}
Similarly, the density variations can be written as
\begin{equation}
\delta n_i(\Vect{x})=\int_{\mu_i}^{\mu_i+\delta\mu_i(\Vect{x})}D_i(\epsilon,\Vect{x})\, d\epsilon \,,
\end{equation}
where $D_i(\epsilon,\Vect{x})$ is the density of states of species $i$. By expanding $D_i(\epsilon,\Vect{x})$ as a power series around $\mu_i$ in the previous equations and dropping higher order terms, we get
\begin{gather}
\delta U_{\text{c,i}}=\int_V \left[ D_i(\mu_i,\Vect{x})\mu_i\delta\mu_i + \left.\frac{\partial}{\partial\epsilon}\left(\epsilon D_i(\epsilon,\Vect{x})\right)\right|_{\mu_i}\!\!\!\!\frac{\delta\mu_i^2}{2}\right] \,d^3x\,,\\
\delta n_i(\Vect{x})= D_i(\mu_i,\Vect{x})\delta\mu_i + \left.\frac{\partial}{\partial\epsilon}D_i(\epsilon,\Vect{x})\right|_{\mu_i}\!\!\!\!\frac{\delta\mu_i^2}{2} \,.
\end{gather}
We can combine the latter equations, again dropping high order terms, to show that 
\begin{equation}
	\delta U_{\text{c,i}}=\int_{V}\left(\mu_i\delta n_i + \frac{\delta n_i \delta \mu_i}{2}\right)\,d^3x \,.
\end{equation}
Using that $\delta n_c\equiv\delta n_p=\delta n_e$, and $\delta\mu_c\equiv\delta\mu_p+\delta\mu_e$ we can write the total amount of energy stored in the perturbation of charged particles as
\begin{align}
\delta U_c&\equiv U_{\text{c,p}}+U_{\text{c,e}} \,,\\
&=\int_{V}\left(\mu_c\delta n_c + \frac{\delta n_c \delta \mu_c}{2}\right)\,d^3x \,.
\end{align}
Finally, as we have a uniform $\mu_c$ and the perturbation $\delta n_c$ conserves the total number of charged particles, the first term is null, leading to 
\begin{equation}
\delta U_c=\int_{V}\frac{\delta n_c \delta \mu_c}{2}\,d^3x \label{eq:dUc}\,.
\end{equation}

Using equations \eqref{eq:dnc} and \eqref{eq:b2def}, it is clear that in an equilibrium state this energy is much smaller than the magnetic energy, 
\begin{equation}
{\delta U_c\over U_B}\sim{\delta n_c\over n_c}\sim b^2 \label{eq:Uc_over_UB}\,.
\end{equation}

Of course, equation \eqref{eq:dUc} is consistent with the loss of magnetic energy due to ambipolar diffusion, which can be computed by taking the derivative of equation \eqref{eq:UB}, and using equation \eqref{eq:induccion0}.
\begin{align}
\frac{dU_B}{dt}&=\frac{1}{4\pi}\int_V d^3x \, \Vect{B}\cdot\frac{\partial\Vect{B}}{\partial t} \,,\\
&=\frac{1}{4\pi}\int_V d^3x \, \Vect{B}\cdot \Curl\left(\Vect{v_A}\times\Vect B\right)\,.
\end{align}
By integrating this expression by parts 
\begin{equation}
\frac{dU_B}{dt}=-\int_{V}\Vect{v_A}\cdot\frac{\Vect{J}\times\Vect{B}}{c}\,d^3x + \frac{1}{4\pi}\int_{\partial V}\Vect{dS}\cdot\left(\Vect{v_A}\times\Vect{B}\right)\times\Vect{B}\,,
\end{equation}
where $\Vect{dS}$ is a surface element normal to the surface $\partial V$ defined by the crust-core interface. We identify this last term as $-dU_{\text{Pol,ext}}/dt$.
Using equation~\eqref{eq:va0}, the first term can be decomposed into the energy loss due to binary collisions
\begin{equation}
\left.\frac{dU_B}{dt}\right|_{\text{Col}}=-\int_V \gamma_{cn}n_c n_n|\Vect{v_A}|^2 \,d^3x \label{eq:dUcol} \,,
\end{equation}
and the variation of energy stored in the particles
\begin{equation}
-\frac{d\delta U_c}{dt}=-\int_V n_c\mu_c\Vect{v_A}\cdot\Grad\left(\frac{\delta\mu_c}{\mu_c}\right) \,d^3x \,.
\end{equation}
Finally, integrating by parts and using equation \eqref{eq:continuidad0} we obtain
\begin{equation}
\frac{d\delta U_c}{dt}=\int_V \delta\mu_c \frac{\partial\delta n_c}{\partial t}\,d^3x \,,
\end{equation}
which is consistent with equation \eqref{eq:dUc}.

\subsection{Magnetic equilibria}
\label{sec:model_magneticequilibria}

As can be seen in equations \eqref{eq:induccion_a}, \eqref{eq:va_a}
and \eqref{eq:continuidad_a}, an equilibrium can only be
achieved when the magnetic force density
$\Vect{f_m}=b^2(\Curl\Vect{B})\times\Vect{B}$ is perfectly balanced by the
degeneracy pressure gradient, yielding $\Vect{v_A}=0$ at all points in the star. Since we are restricted to axial
symmetry, there cannot exist a pressure gradient in the azimuthal direction to
balance the magnetic force, so that in equilibrium we must have $f_{m\,\phi} \propto
|\Grad\alpha\times\Grad\beta| = 0$. Therefore, $\Grad\alpha\parallel\Grad\beta$, which in turn implies that, at least locally, there is a one-to-one relation between the functions $\alpha(r,\theta)$ and $\beta(r,\theta)$. One important consequence of this is that, since $\beta=0$ outside the neutron star core, and poloidal field lines correspond to lines of constant $\alpha$, $\beta$ (and thus the toroidal field) can be non-zero only along the field lines that close within the core. The geometric reason for this is that open field lines can ``unwind'', eliminating their toroidal component and thus reducing their magnetic energy, whereas closed field lines cannot do this because of topological constraints.

Writing, for definiteness, $\beta=\beta(\alpha)$ (although this function may not be single-valued), we have $\Grad\beta=\beta'\Grad\alpha$, where $\beta'=d\beta/d\alpha$.
The poloidal part of the ambipolar velocity (equation \ref{eq:va_AB}) then reduces to
\begin{equation}
	\frac{b^2}{r^2\sin^2\theta}\left(\Sh\alpha+\beta\beta'\right)\Grad\alpha + \Grad\chi = 0 \label{eq:GS0} \,.
\end{equation}
This implies $\Grad\alpha\parallel\Grad\chi$, and hence $\chi=\chi(\alpha)$, leading to
\begin{equation}
	\Sh\alpha+\beta\beta' + \frac{r^2\sin^2\theta}{b^2}\chi' = 0 \,.\label{eq:GS}
\end{equation}
This so-called ``Grad-Shafranov equation''
\citep{Grad1958,Shafranov1966}, which relates the functions $\alpha(r,\theta)$, $\beta(\alpha)$, and $\chi(\alpha)$; provides magneto-hydrostatic
equilibrium configurations for a given charged particle
density (and chemical potential) profile \citep[e.g.,][]{Armaza2015}. However, their stability is not guaranteed by the GS equation. Hence, in this manuscript we
will analyze the stability (in axial symmetry) of certain equilibrium configurations, and test if the asymptotic solutions of our simulations do converge to equilibrium states for which $\beta$ and $\chi$ are functions of $\alpha$ and which satisfy equation \eqref{eq:GS}.

In order to evolve the magnetic field in the NS core, a numerical
code has been developed. The code evolves the set of equations \eqref{eq:continuidad_a}, \eqref{eq:evolucionalfa}, and \eqref{eq:evolucionbeta}, with the vector fields $\Vect{v_A}$ and $\Vect B$ given by eqs. \eqref{eq:va_a} and \eqref{eq:BAlfaBeta}, respectively. 
This is done by discretizing the values of the variables over a staggered polar grid composed of $N_r$ points inhomogeneously distributed in the radial direction inside the core and $N_\theta$ points equally spaced in the polar direction, while the external multipolar expansion (see equation \ref{multipolos}) is truncated to the first $N_\text{Exp}$ terms.
Numerical computation is done conservatively for the evolution of the toroidal magnetic field and the density perturbation of the charged particles.
The time derivative of the poloidal potential is computed using finite difference, and the system is evolved to second-order accuracy in time.
Details on the numerical code can be seen in Appendix \ref{app:code}, while some of the numerical tests performed to ensure its functionality can be found in Appendix \ref{app:tests}.

\section{Results}
\label{sec:results}

\subsection{Characteristic evolutionary time-scales}
\label{sec:results_timescales}

In order to check whether the time-scales $t_c$ and $t_B$ (equations \ref{eq:tc} and \ref{eq:tb}, respectively) are in agreement with our numerical code, we have performed simulations taking combinations of Ohmic modes as initial conditions. These are normal mode solutions to equation~\eqref{eq:induccion} when we set the Hall, ambipolar, and battery terms to zero, namely
\begin{equation}
	\alpha_{nl}=A_{nl}\, rj_l(k_{nl} r)P_l^1(\cos\theta)\sin\theta\,, \label{eq:ohm}
\end{equation}
where $j_l(x)$ is the spherical Bessel function of order $l$. The values of $k_{nl}$ are set by the boundary conditions, where $n$ is the
radial index of the mode. The solutions of this equation for the toroidal field, $\beta_{nl}$, involve the same radial and angular functions, but the values of $k_{nl}$ are different because of the different boundary conditions imposed for the poloidal and toroidal fields. For the poloidal magnetic field we impose continuity of $\alpha$ and its radial derivative with the multipolar expansion outside, and for the toroidal magnetic field we impose $B_\phi=0$ at the boundary, hence $\beta=0$. The amplitude $A_{nl}$ is set by requiring that the volume-average of $B^2$ over the interior of the star is equal to one. The values of these constants are shown in Table \ref{tab:poloidal} and Table \ref{tab:toroidal}, respectively, for the first poloidal and toroidal Ohmic modes.

\begin{table}
	\caption{Values of $A_{nl}$ and $k_{nl}$ for the first poloidal Ohmic modes (functions $\alpha_{nl}$; see equation \ref{eq:ohm}).}  
	\label{tab:poloidal}		
	\begin{tabular}{cccc}
		\hline
		& $n=1$ & $n=2$ & $n=3$ \\
		\hline
		$A_{n1}$	& 1.1199 & 1.0263 & 1.0115 \\
		$k_{n1}$	& $\pi$ & $2\pi$ & $3\pi$ \\
		\hline
		$A_{n2}$	& 0.8527 & 0.7781 & 0.7614 \\
		$k_{n2}$	& 4.4934 & 7.7253 & 10.9041 \\
		\hline
		$A_{n3}$	& 0.722 & 0.6596 & 0.6427 \\
		$k_{n3}$	& 5.7635 & 9.095 & 12.3229 \\
		\hline
		$A_{n4}$	& 0.6406 & 0.5857 & 0.569 \\
		$k_{n4}$	& 6.9879 & 10.4171 & 13.698 \\
		\hline
		$A_{n5}$	& 0.5836 & 0.5339 & 0.5175 \\
		$k_{n5}$	& 8.1826 & 11.7049 & 15.0397 \\
		\hline
	\end{tabular}
\end{table}

\begin{table}
	\caption{Values of $A_{nl}$ and $k_{nl}$ for the first toroidal Ohmic modes (functions $\beta_{nl}$; see equation \ref{eq:ohm}).}
	\label{tab:toroidal}		
	\begin{tabular}{cccc}
		\hline
		& $n=1$ & $n=2$ & $n=3$ \\
		\hline
		$A_{n1}$	& 4.6033 & 7.7897 & 10.9499 \\
		$k_{n1}$	& 4.4934 & 7.7253 & 10.9041 \\
		\hline
		$A_{n2}$	& 4.5024 & 6.9052 & 9.277 \\
		$k_{n2}$	& 5.7635 & 9.095 &  12.3229 \\
		\hline
		$A_{n3}$	& 4.6603 & 6.686 & 8.6832 \\
		$k_{n3}$	& 6.9879 & 10.4171 & 13.698 \\
		\hline
		$A_{n4}$	& 4.8793 & 6.6651 & 8.4289 \\
		$k_{n4}$	& 8.1826 & 11.7049 & 15.0397 \\
		\hline
		$A_{n5}$	& 5.1170 & 6.7293 & 8.3283 \\
		$k_{n5}$	& 9.3558 & 12.9665 & 16.3547\\
		\hline
	\end{tabular}
\end{table}

\begin{figure}
	\centering
	\includegraphics[width=\linewidth]{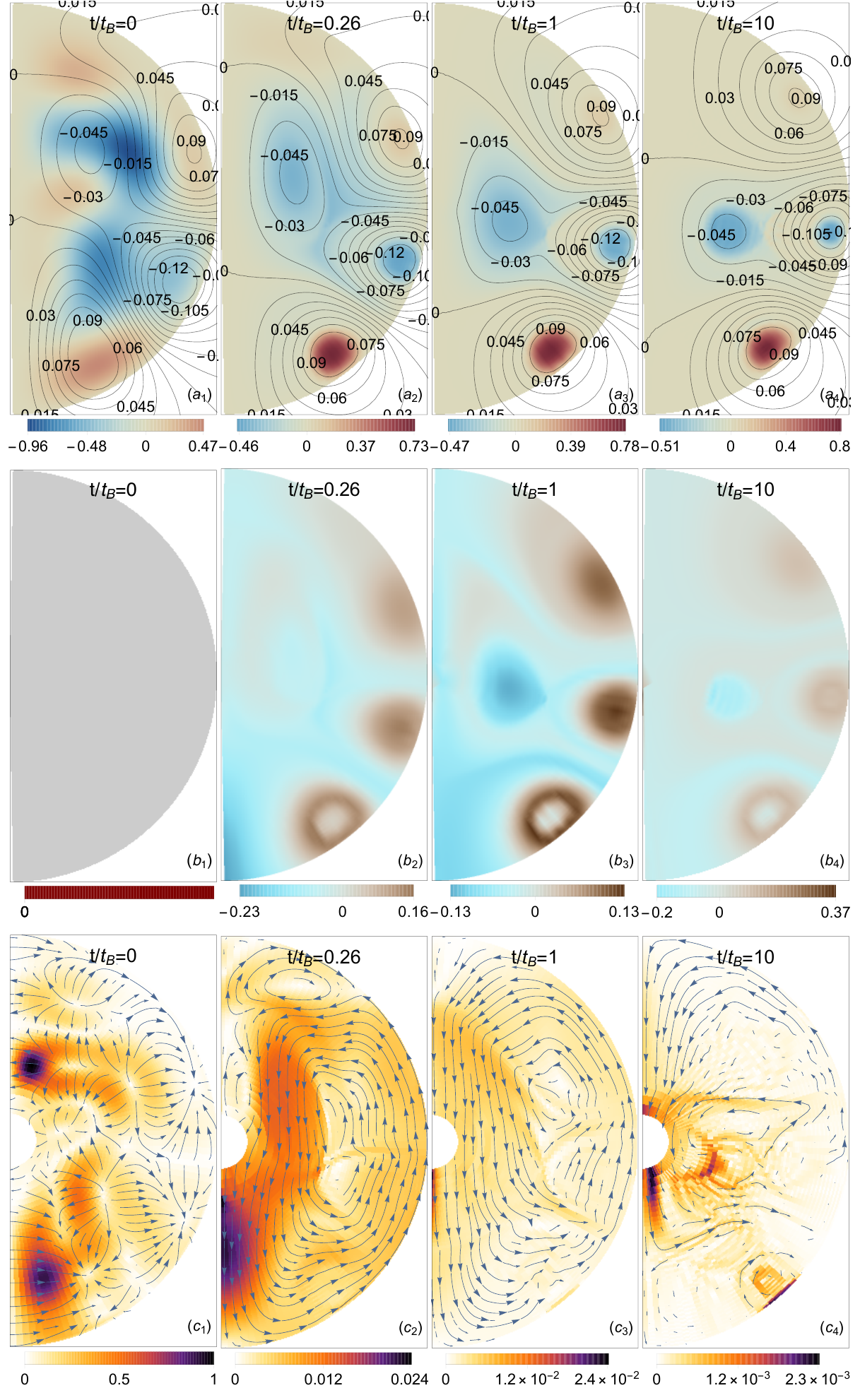}
	\caption{Configuration of the magnetic field at different times $t/t_B$ (top panels), where lines represent the poloidal magnetic field and the colours the toroidal potential. The initial magnetic field is constructed with a combination of $\alpha = (\alpha_{21} + 3\alpha_{22} - 2\alpha_{13} + 4\alpha_{14} - 3\alpha_{15})/\sqrt{78}$ and $\beta = (2\beta_{11} - \beta_{31} + \beta_{13} + \beta_{34} - \beta_{15})/4$, where both components have the same initial magnetic energy. The middle panels show the evolution of the particle density perturbation $\chi$ in time, which is initially taken to be null at all points. The bottom panels show the poloidal component of the ambipolar diffusion velocity $\Vect{v_A}$, where arrows represent its direction and colours its magnitude normalized to its maximum initial value. The simulation is performed for $b^2=0.31$ on a grid of $N_r,N_\theta,N_{\text{Exp}}=60,91,27$, respectively. }
	\label{fig:pol_tor_raro}
\end{figure}

\begin{figure}
	\centering
	\includegraphics[width=\linewidth]{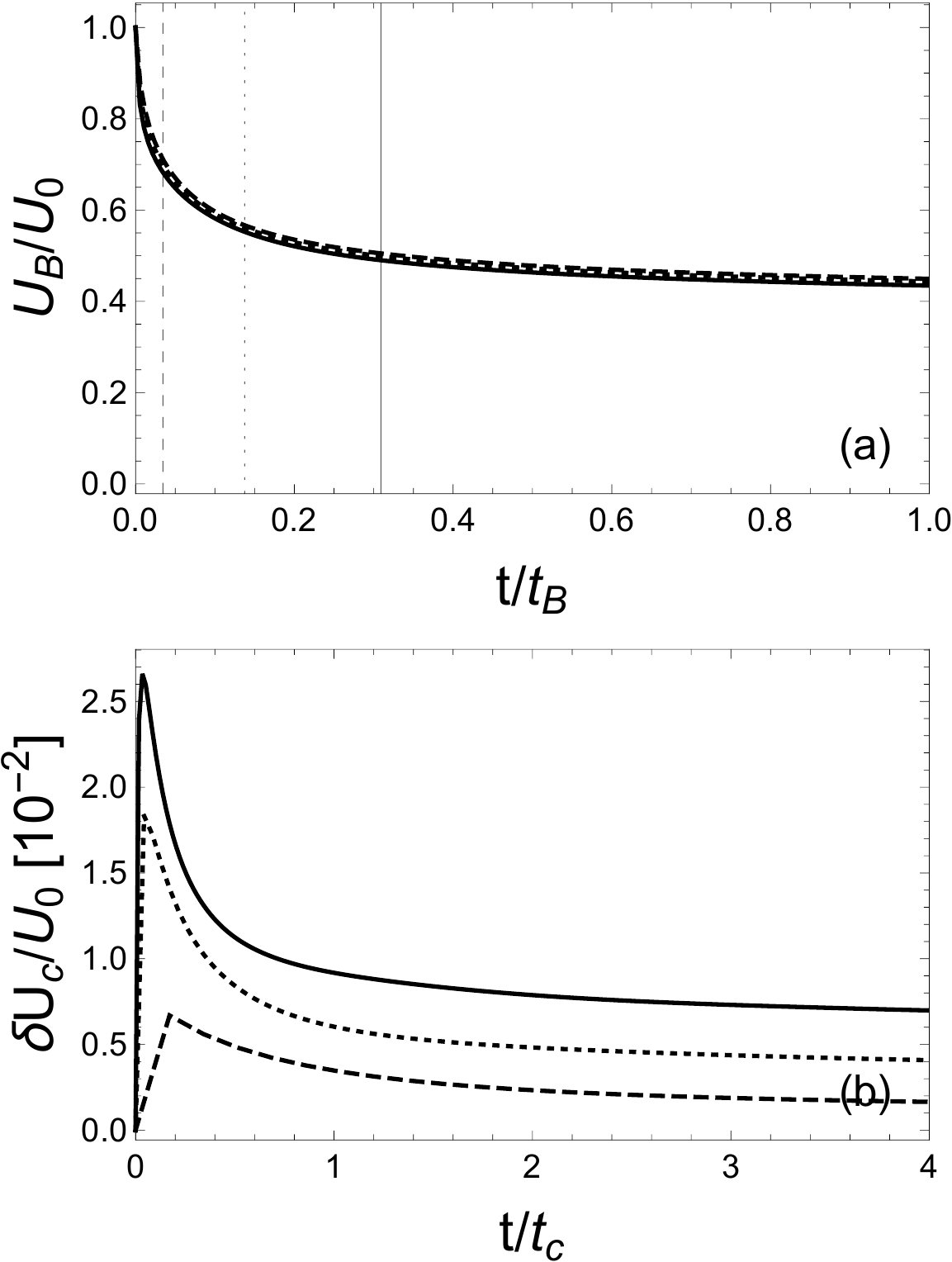}
	\caption{(a) Time evolution of the magnetic energy in the simulation described in Fig.~\ref{fig:pol_tor_raro}, but using $b^2=0.03$(dashed), $b^2=0.14$(dotted), $b^2=0.31$(continuous line). The vertical lines show the respective values of $t_c=b^2 t_B$. For each case time is normalized in units of $t_B$. (b) Particle energy for the same simulations, with time normalized to $t_c$.}
	\label{fig:compb_raro}
\end{figure}

The initial configuration of one of our simulations and some snapshots of its evolution in time can be seen in Fig.~\ref{fig:pol_tor_raro}, where we observe that, for this magnetic field configuration, $\Vect{v_A}$ initially has a strong irrotational component, which pushes the magnetic field and the charged particles from regions of positive to negative $\Div\Vect{v_A}$.  We see that, shortly after the simulations starts, this excess of charged particles generates a pressure gradient that cancels the irrotational component of the velocity, leaving the velocity field mostly solenoidal, as can be seen at $t/t_B=0.26$. The velocity field keeps moving both the charged particles and the magnetic field across larger distances, but eventually, after a few $t_B$, the system reaches a state in which the magnetic force density is perfectly balanced by the gradient of the degeneracy pressure, and ambipolar diffusion is effectively suppressed, so that no relevant further evolution is seen. Furthermore, Fig.~\ref{fig:pol_tor_raro} shows that at $t/t_B=10$ the magnitude of $\Vect{v_A}$ has decreased by a factor
$10^3$ from its initial value. Hence, the magnetic field reaches an equilibrium state.

Fig.~\ref{fig:compb_raro}(a) shows that the evolution of the magnetic energy as a function of $t/t_B$ is nearly identical for three quite different values of the parameter $b$ characterizing the magnetic field strength, supporting the expectation that the relevant time-scale for its evolution is proportional to $t_B$, although its actual value is substantially smaller, $\lesssim 0.1 t_B$. Fig.~\ref{fig:compb_raro}(b) also shows how within a fraction of $t_c$, roughly the same for the three cases, the irrotational component of the magnetic force is balanced by the degeneracy pressure gradient leading to a peak of particle energy, whose magnitude is small compared to the magnetic energy, in fact about one order of magnitude smaller than the estimate of equation \eqref{eq:Uc_over_UB}. Afterwards, it decays much more slowly, probably because of the readjustment of the magnetic field on a time-scale $\propto t_B$, eventually reaching a plateau corresponding to the final equilibrium state.

\subsection{Equilibrium configurations}
\label{sec:results_equilibriumconfigurations}

In the simulation described above, it can be seen how within a few $t_B$ the star reaches a state in which ambipolar diffusion is null.  One can also see that, although the initial configuration was deliberately constructed so that $\beta\neq 0$ also on open poloidal field lines, in the final equilibrium state $\beta$ is non-zero only on field lines that close within the core, as expected from the analysis in section \ref{sec:model_magneticequilibria}. It is interesting to explore if this equilibrium is indeed a solution of the GS equation~\eqref{eq:GS}. 

\begin{figure}
	\centering
	\includegraphics[width=\linewidth]{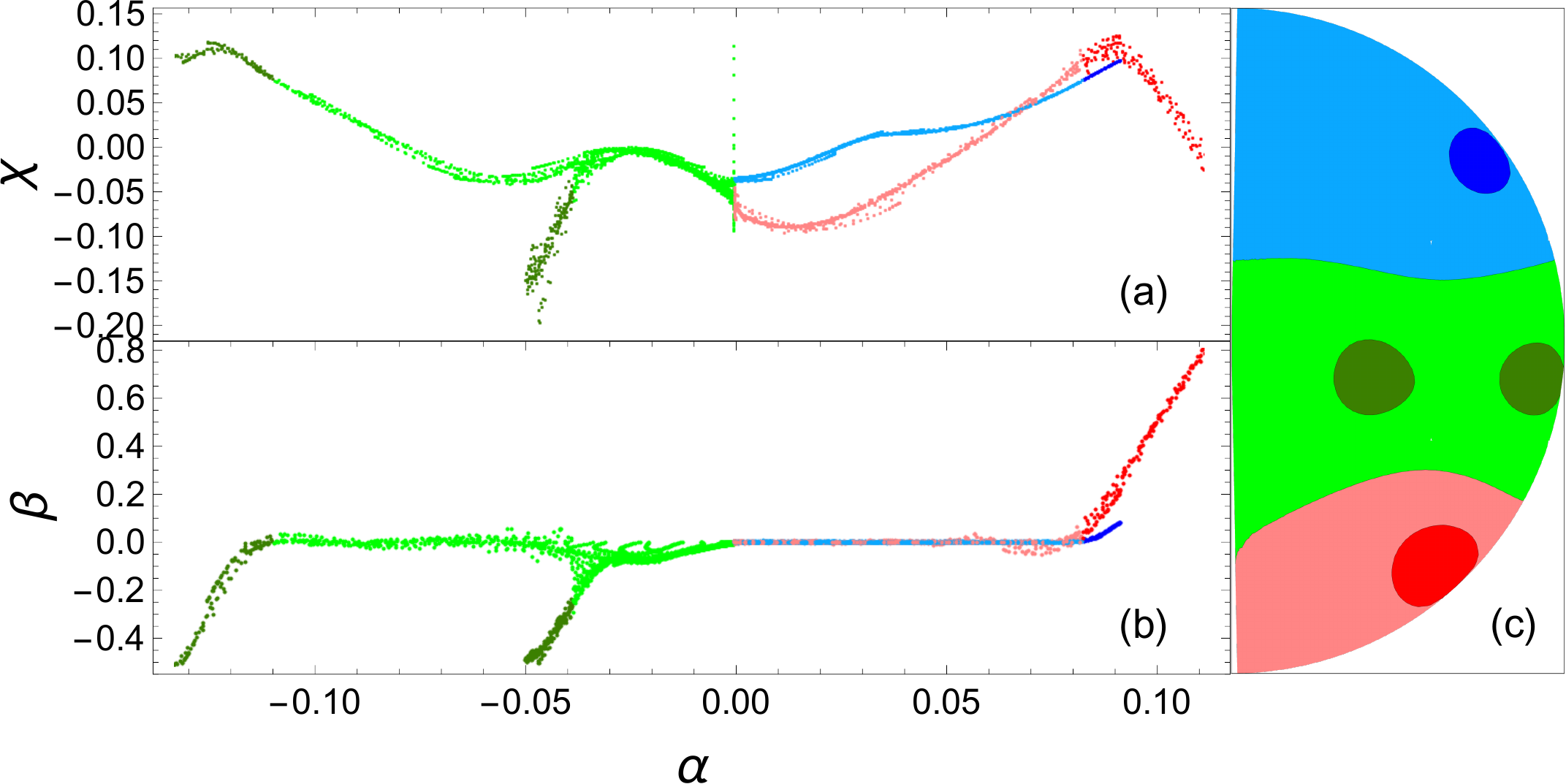}
	\caption{(a) $\chi$ vs. $\alpha$, and (b) $\beta$ vs. $\alpha$ at $t/t_B=10$. In both panels, the colouring of each point corresponds to its colour in panel (c). The initial magnetic configuration and the simulation parameters are described in Fig.~\ref{fig:pol_tor_raro}.}
	\label{fig:pol_tor_dnc_vs_alfa_raro}
\end{figure}

Equation~\eqref{eq:GS} forces us to have $\beta=\beta(\alpha)$, and $\chi=\chi(\alpha)$ at equilibrium.
Fig.~\ref{fig:pol_tor_dnc_vs_alfa_raro}(a) and (b) show these relations at $t/t_B=10$ for the simulation described in Fig.~\ref{fig:pol_tor_raro}.
Here we see how in this case $\chi(\alpha)$ and $\beta(\alpha)$ are not single-valued, but, since at least locally one of the variables can be written as a function of the other, the equilibrium condition holds.
In both plots we see four different branches, which correspond to the four different regions where the toroidal magnetic field is enclosed.

In order to assess whether the GS equation \eqref{eq:GS} is satisfied, we evaluate the quantity
\begin{equation}
	\Gamma=\frac1V\int_V \frac{d^3x}{\max\left|\Sh\alpha\right|}\left|\Sh\alpha+\beta\beta' + \frac{r^2\sin^2\theta\chi'} {b^2} \right|\,.
\end{equation}
We get $\Gamma(t=0)=0.42$ and $\Gamma(t=t_B)=6\times10^{-3}$. Therefore, at $t=t_B$ the core is much closer to be a solution of GS than initially.

\begin{figure}
	\centering
	\includegraphics[width=0.66\linewidth]{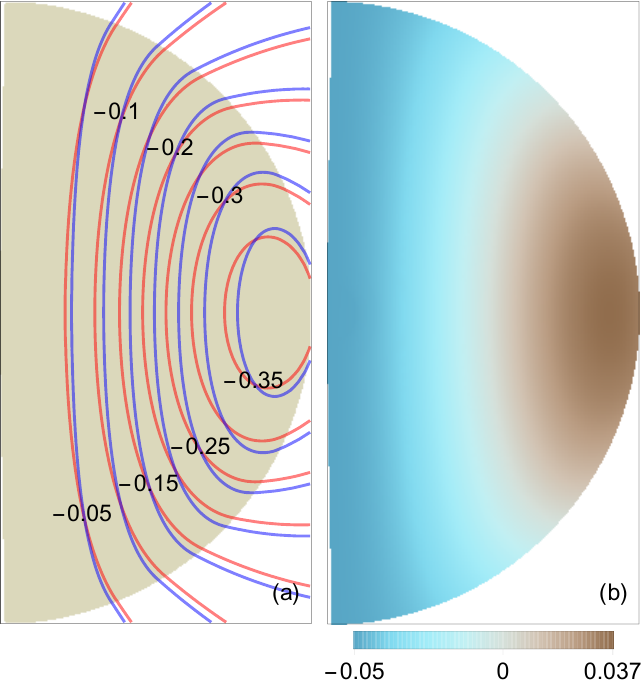}
	\caption{Simulation discussed in Sec.~\ref{sec:results_equilibriumconfigurations}. (a) Red lines represent equipotentials of $\alpha$ (labeled by the values of this variable), which correspond to poloidal magnetic field lines of the initial magnetic field configuration used in the simulation: A purely poloidal Ohmic mode $\alpha_{11}$. The initial values of $\beta$ and $\chi$ are null at all points. The simulation was performed for $b^2=0.03$ using a grid of $N_r,N_\theta,N_{\text{Exp}}=60,91,27$, respectively. Blue lines represent the same equipotentials of $\alpha$ at $t/t_B=1$, after equilibrium has been reached. (b) $\chi\equiv\delta n_c/n_{c}$ at $t/t_B=1$, where colours represent its magnitude.}
	\label{fig:alfa_puro}
\end{figure}

\begin{figure}
	\centering
	\includegraphics[width=\linewidth]{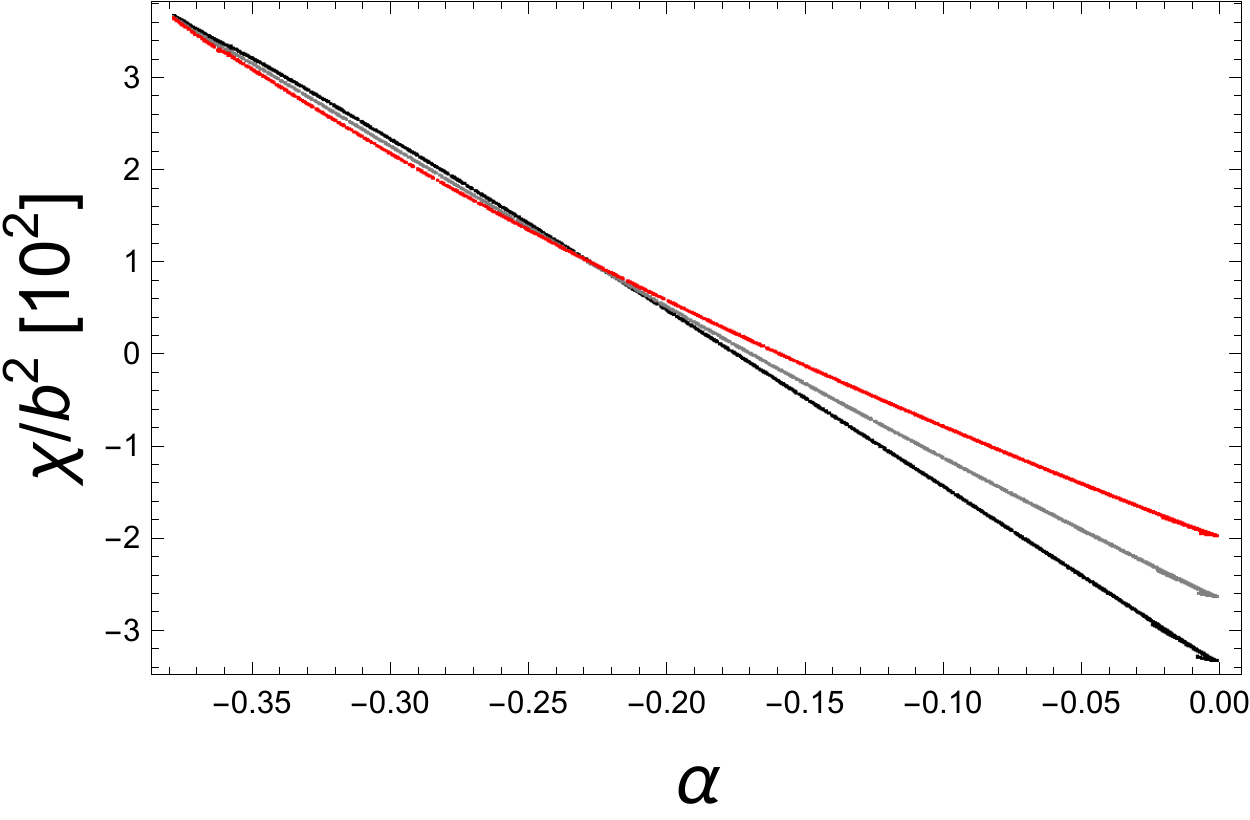}
	\caption{Scatter plot of $\alpha$ vs. $\chi/b^2$ showing all of the grid points at $t/t_B=1$, after equilibrium has been reached, with magnetic field strengths $b^2=0.03$(black), $b^2=0.14$(gray) and, $b^2=0.31$(red), respectively. The simulation parameters and the initial condition are described in Fig.~\ref{fig:alfa_puro}.}
	\label{fig:alfa_vs_dnc}
\end{figure}

We have also performed a numerical simulation taking as initial condition the purely poloidal Ohmic mode $\alpha_{11}=-A_{11} rj_1(\pi r)\sin^2\theta$, where $A_{11}=1.12$, which can be seen in Fig.~\ref{fig:alfa_puro}(a) (red lines). We let the system evolve up to $t/t_B=1$, where we stop seeing relevant variations in the configuration of the field. The final configuration of the magnetic field, as well as the final configuration of $\chi$ can be seen in Fig.~\ref{fig:alfa_puro}(a) (blue lines) and (b), respectively. Fig.~\ref{fig:alfa_vs_dnc} shows that the relation $\chi=\chi(\alpha)$ is satisfied globally and that the dependence on $\alpha$ is roughly linear, although the particular shape is not relevant, as long as there is a well-defined dependence of $\chi$ on $\alpha$. This plot also shows that $\chi$ scales properly with the magnetic field intensity, as the magnitude of $\chi/b^2$ is roughly the same for the three values of $b^2$ used.

As initially we have $\chi=0$, and $\beta=0$, the maximum value of $\Gamma$ is 0.56, but after $t/t_B=0.3$, its value is less than $3\times10^{-3}$, suggesting that the system is converging to an equilibrium satisfying the GS equation \eqref{eq:GS}.
Hence, ambipolar diffusion is able to take the system to a purely poloidal equilibrium configuration that seems to be stable, although it is likely that this axially symmetric equilibrium is not stable under non-axially symmetric perturbations.

In the case of a purely toroidal initial field, similar results are expected. Although these configurations are also known to be unstable \citep{Braithwaite2009}, they are simple cases in which we can test if the code is working properly as well as getting some interesting conclusions.

\begin{figure}
	\centering
	\includegraphics[width=\linewidth]{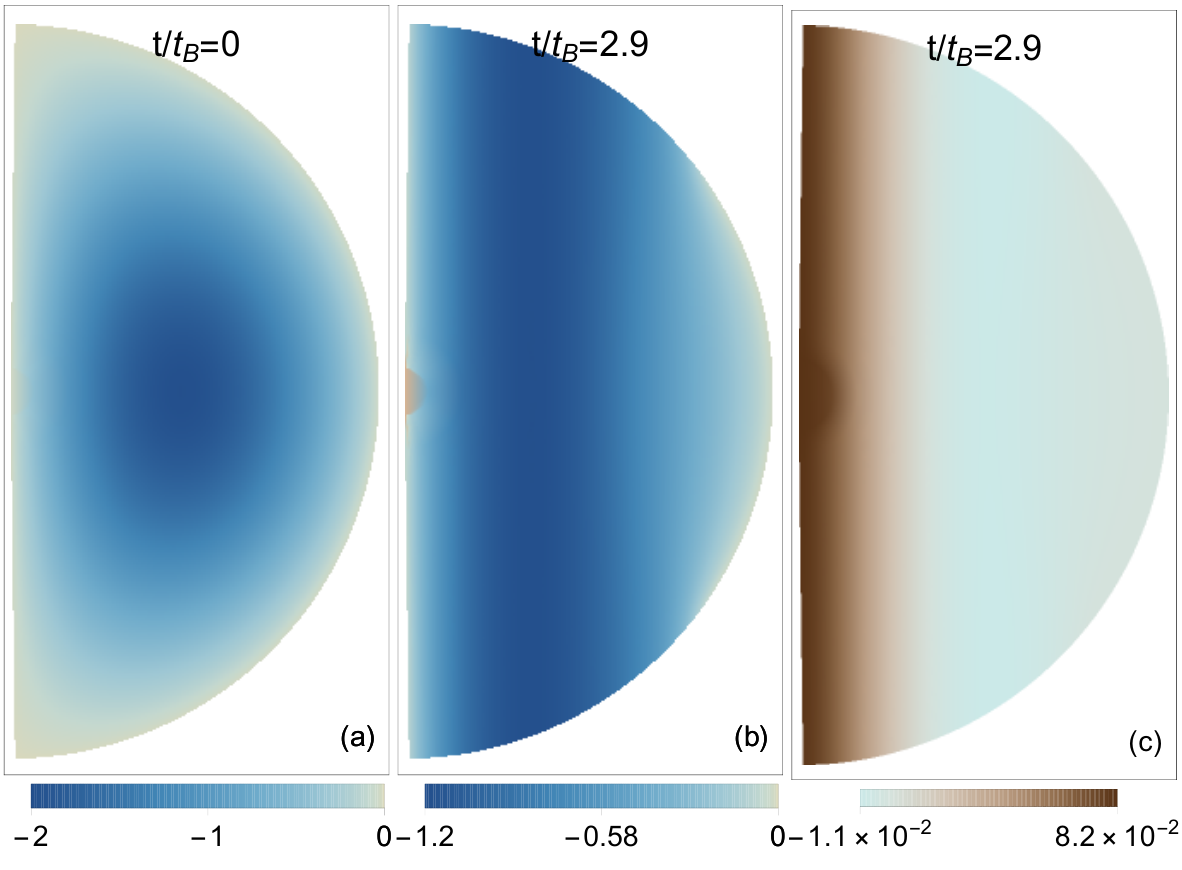}
	\caption{(a) Initial toroidal magnetic field configuration used in the simulation: A purely toroidal Ohmic mode $\beta_{11}$. The initial value of $\alpha$ and $\chi$ is null at all points. The simulation was performed for $b^2=0.03$ using a grid of $N_r,N_\theta,N_{\text{Exp}}=60,121,24$, respectively. (b) Toroidal magnetic field and, (c) $\chi$ at $t/t_B=2.9$, after equilibrium has been reached.}
	\label{fig:beta_puro}
\end{figure}

\begin{figure}
	\centering
	\includegraphics[width=\linewidth]{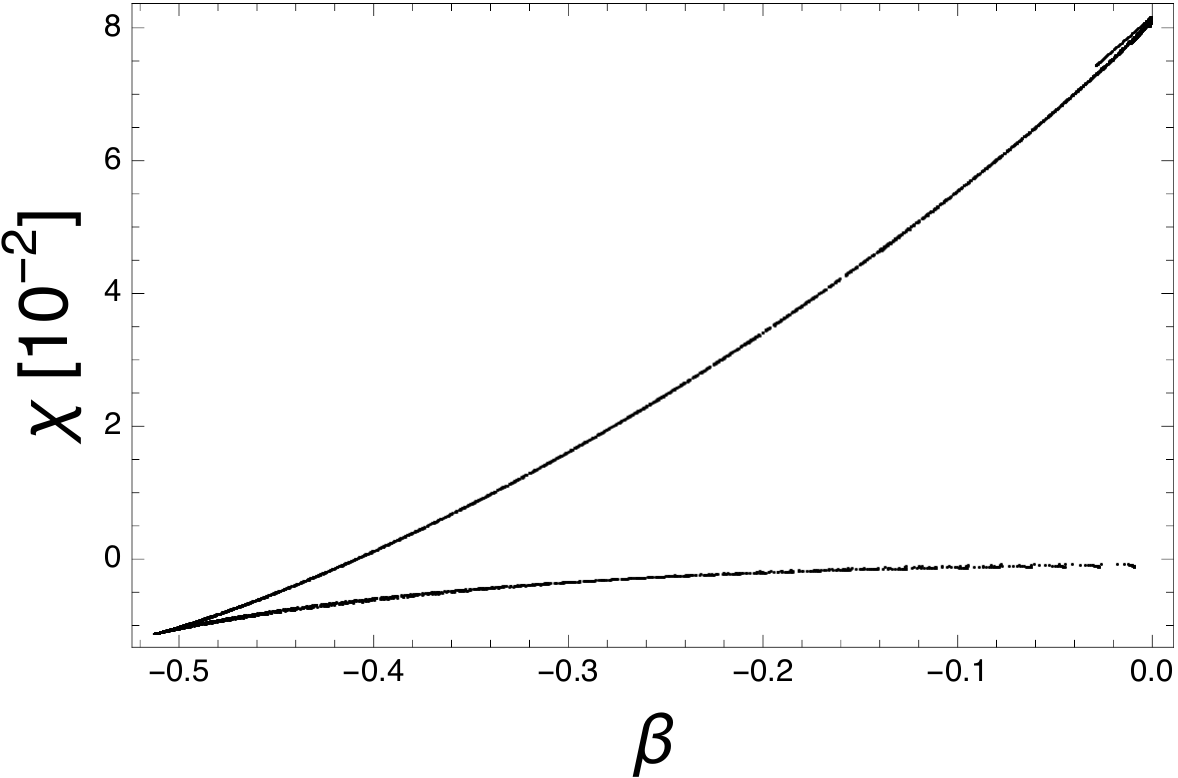}
	\caption{Relation between $\chi$ and $\beta$ at $t/t_B=2.9$, using the data plotted in Fig.~\ref{fig:beta_puro}.}
	\label{fig:beta_vs}
\end{figure}

We start with a purely toroidal Ohmic mode $\beta_{11}$, which is shown in Fig.~\ref{fig:beta_puro}(a). After $t/t_B=2.9$ the system reaches an equilibrium state.
As we can see in Fig. $\ref{fig:beta_puro}$(b) and (c), the system converges to a configuration in which both $\beta$ and $\chi$ are functions of the cylindrical radius $\rho=r\sin\theta$.
This can be understood from equation~\eqref{eq:va_AB} by setting $\alpha=0$ and imposing $\Vect{v_A}=0$.
The equation reduces to
\begin{equation}
\frac{b^2}{r^2\sin^2\theta}\beta\Grad\beta + \Grad\chi = 0 \,,
\end{equation}
thus forcing a dependence $\chi=\chi(\beta)$. Thus, equilibrium configurations must satisfy
\begin{equation}
\frac{b^2}{\rho^2} = -\frac{1}{\beta}\frac{d\chi}{d\beta} \,, \label{eq:GSbeta}
\end{equation}
which tells us that both $\beta$ and $\chi$ must be functions of $\rho$. Therefore, for toroidal fields, any equilibrium must have cylindrical symmetry.
Fig.~\ref{fig:beta_vs} shows the relation between $\beta$ and $\chi$ from the simulation. Again, $\chi(\beta)$ is not single-valued, but since at least locally one of the variables can be written as a function of the other, the equilibrium condition holds.

\begin{figure}
	\centering
	\includegraphics[width=\linewidth]{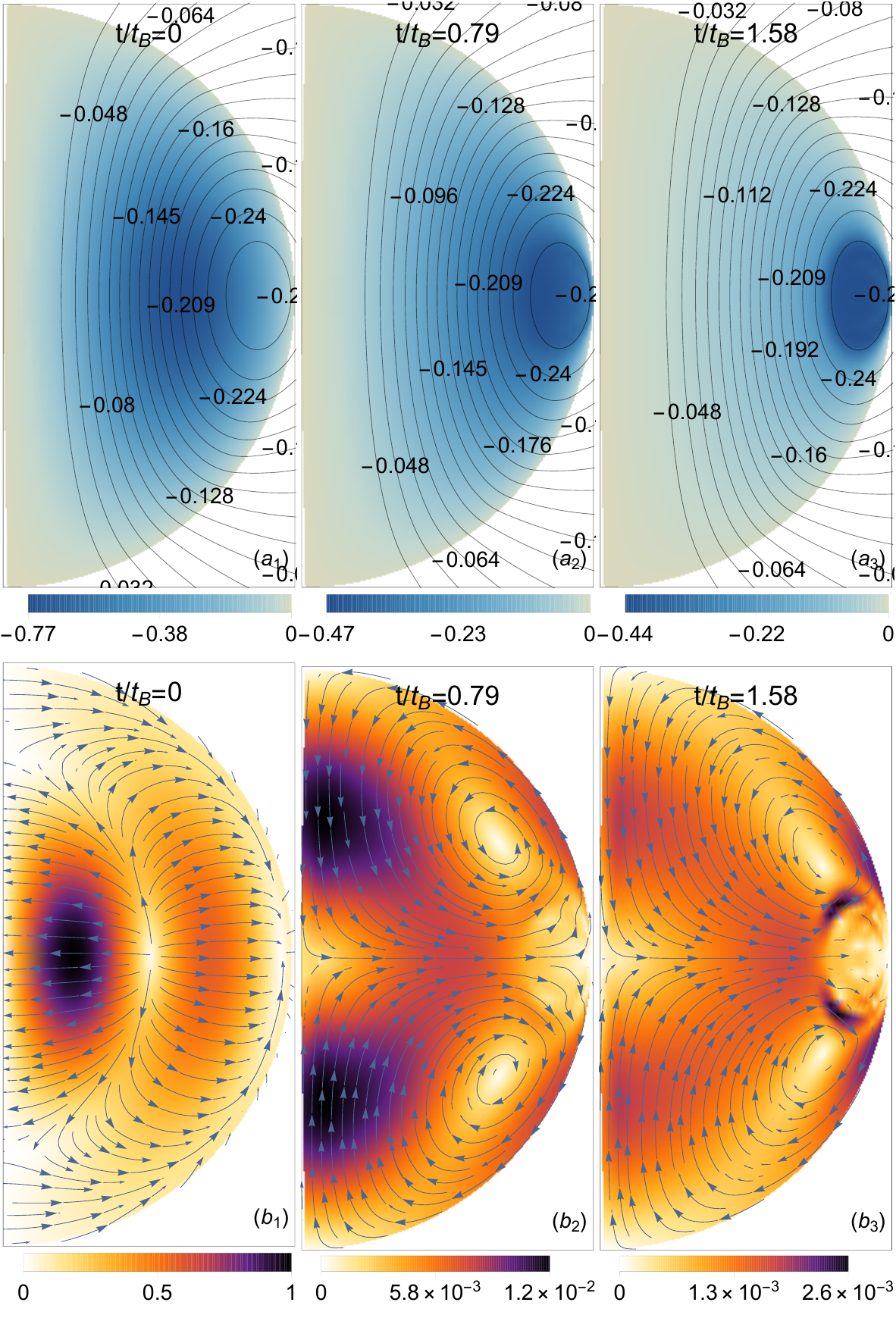}
	\caption{
		Evolution of a magnetic field whose initial condition is a combination of the fundamental poloidal and toroidal Ohmic modes ($\alpha_{11}/\sqrt2$ and $\beta_{11}/\sqrt2$) with equal energies. In the upper panels the lines represent the poloidal magnetic field and the colours the toroidal potential. In the lower panels, arrows represent the direction of the poloidal component of $\Vect{v_A}$, and the colour represents its magnitude normalized to its maximum initial value. The simulation was performed for $b^2=0.31$ using a grid of $N_r,N_\theta,N_{\text{Exp}}=60,91,27$, respectively.}
	\label{fig:a11b11}
\end{figure}

 \begin{figure}
 	\centering
	\includegraphics[width=\linewidth]{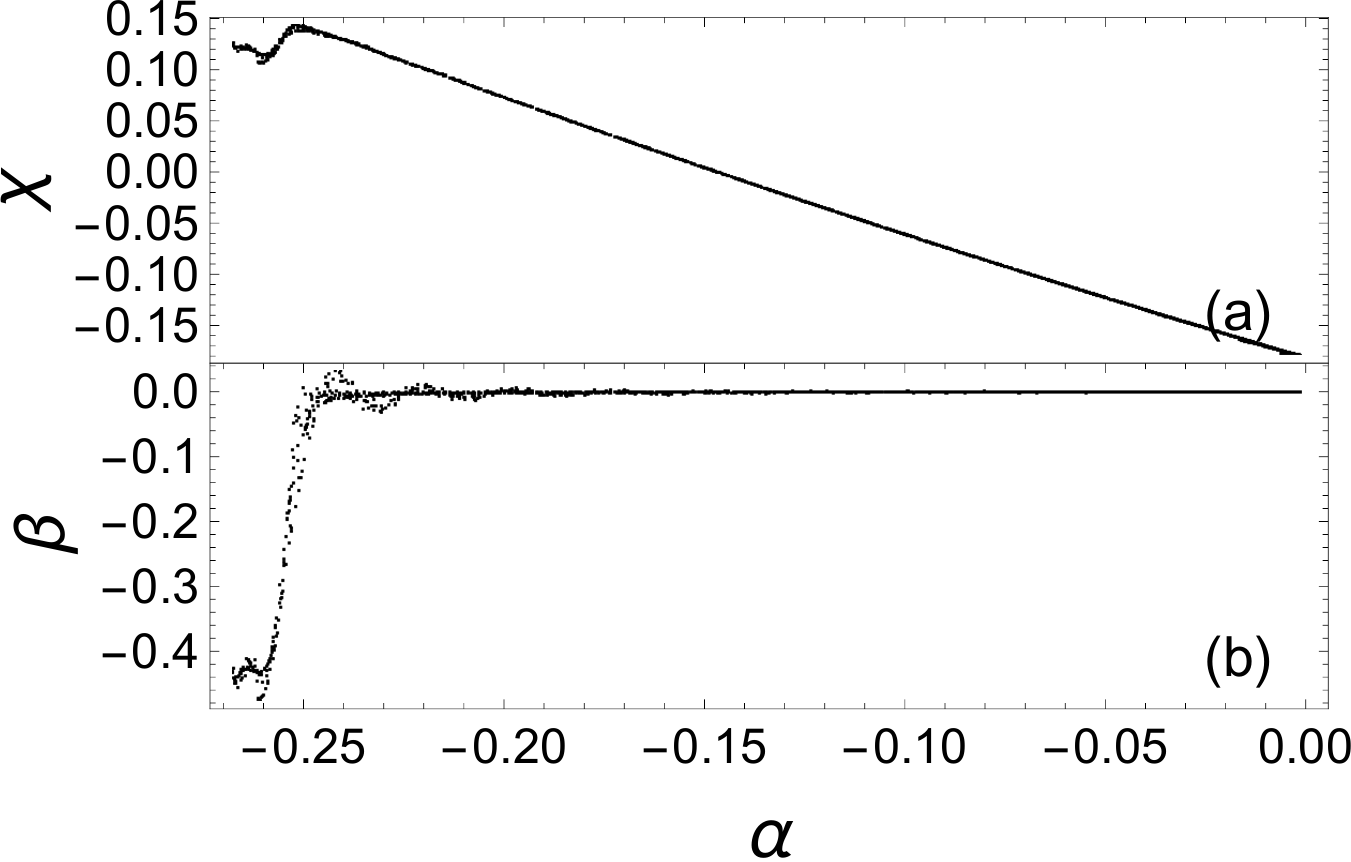}
 	\caption{(a) $\chi$ vs. $\alpha$, and (b) $\beta$ vs. $\alpha$ at $t/t_B=17.4$ for the simulation shown in Fig.~\ref{fig:a11b11}.}
 	\label{fig:pol_tor_dnc_vs_alfa}
 \end{figure}

In the simulation shown in Fig.~\ref{fig:pol_tor_raro}, we see that the magnetic field tends to form structures in which the poloidal magnetic field extends throughout the entire star and to the exterior, while the toroidal field is confined to torus-shaped regions defined by the poloidal field lines that close inside the star.
To test if it is possible to get similar configurations from other initial conditions we perform a simulation taking an initial magnetic field constructed as a superposition of the fundamental poloidal and toroidal Ohmic modes, both having the same initial energy.
Fig.~\ref{fig:a11b11} shows that, as in the simulation shown in Fig.~\ref{fig:pol_tor_raro}, the field evolves towards an equilibrium state in which the magnetic field lines with a substantial toroidal component close inside the star.

A comparison of Figs.~\ref{fig:pol_tor_dnc_vs_alfa_raro}(a),  \ref{fig:alfa_vs_dnc}, and \ref{fig:pol_tor_dnc_vs_alfa}(a) shows that the relation between $\chi$ and $\alpha$ is quite similar for this configuration compared to the ones studied previously, with an essentially linear behaviour in the region with open poloidal field lines and $\beta=0$ and a somewhat reduced value of $\chi$ in the torus where $\beta\neq 0$.
Fig.~\ref{fig:pol_tor_dnc_vs_alfa}(b) shows the relation between $\beta$ and $\alpha$, which is similar to the one shown in Fig.~\ref{fig:pol_tor_dnc_vs_alfa_raro}(b), suggesting that both equilibrium configurations are essentially the same.

As can be seen in Fig.~\ref{fig:a11b11}, initially the ambipolar velocity is mainly driven by the magnetic force and has a big region close to the symmetry axis in which a strong velocity is pointing towards the axis. However, the pressure gradient generated by this motion quickly balances this force, and by $t/t_B=0.79$ (Fig.~\ref{fig:a11b11}(b$_2$)) we see how the poloidal component of $\Vect{v_A}$ is mainly solenoidal, pushing the particles and the magnetic field towards the region in which the magnetic field lines close inside the star, forming a ``twisted torus''.
	
This gives us some insight on what conditions are to be met by
the initial conditions to get similar equilibrium configurations.
Since in this model the magnetic field is fixed to the charged
particles, and the particles cannot leave the star, the closed poloidal field lines
inside of the star will remain inside. Also, since we do not have
magnetic reconnection (as long as numerical dissipation can be ignored), these loops cannot split nor merge, therefore,
the number of ``twisted tori'' found in an equilibrium configuration
have to be the same as the number of different regions of closed
magnetic field loops inside of the star in the initial condition. This
is illustrated by Fig.~\ref{fig:pol_tor_raro}, where the initial poloidal field has four regions in which the magnetic field lines close inside of the star, and thus four ``twisted tori'' are present in the final equilibrium state. In the regions of open poloidal field lines, on the other hand, the field can ``unwind'' through a toroidal component of $\Vect{v_A}$, thus leaving $\beta=0$ in the final state (see section \ref{sec:model_axially_symmetric_fields}).

\subsection{Stability of a simple equilibrium found by solving the GS equation}
\label{sec:results_stabilityofequilibria}

\begin{figure}
	\centering
	\includegraphics[width=\linewidth]{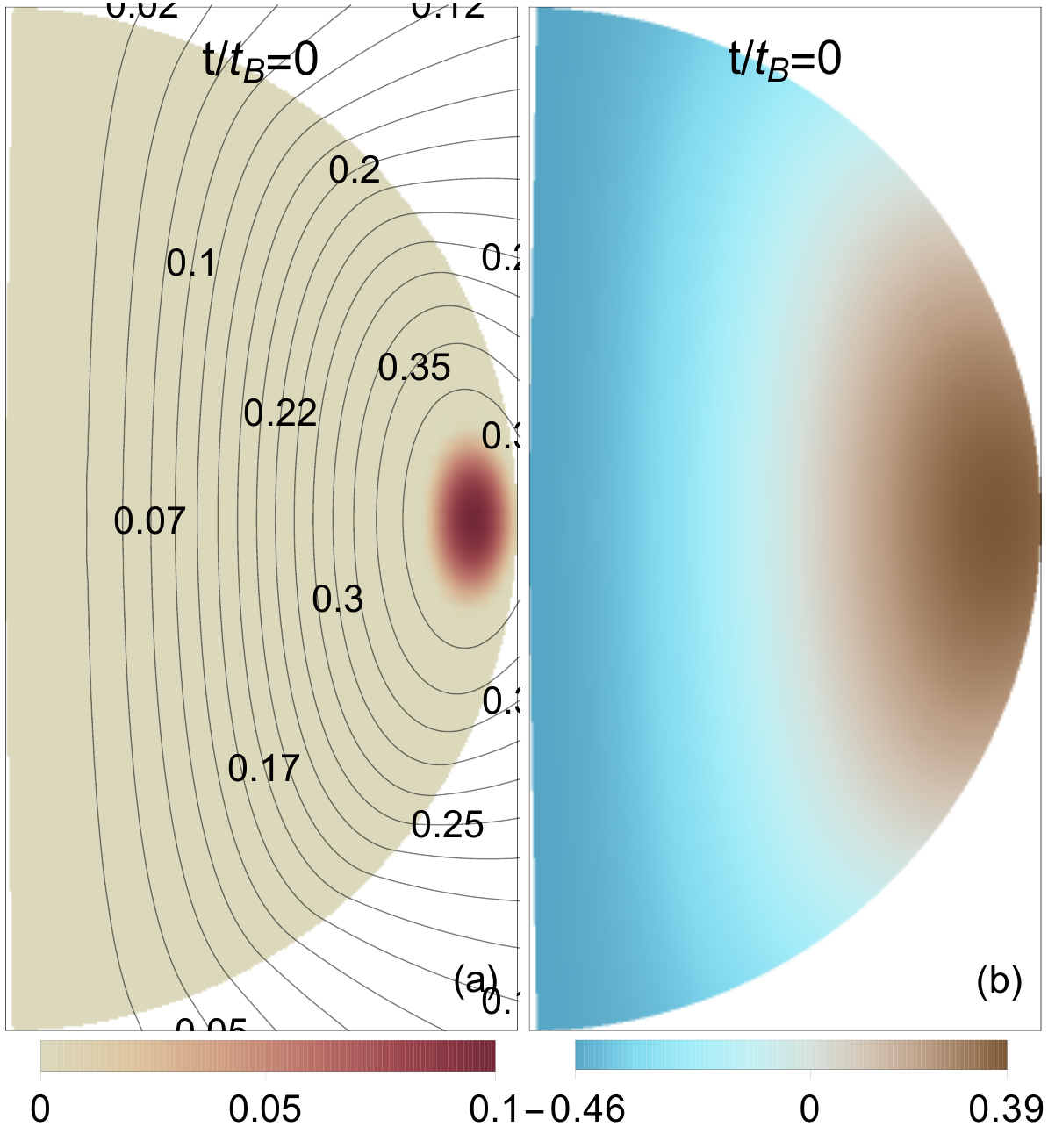}
	\caption{(a) Initial configuration: Magnetic field configuration described in equation~\eqref{eq:twisted} where $\alpha(r,\theta)$ is a numerical solution of equation~\eqref{eq:GS}, taking $s=10$. Lines represent equipotentials of $\alpha$, which correspond to magnetic field lines. Colours represent the magnitude of the toroidal potential $\beta$. (b) $\chi\equiv\delta n_c/n_{c}$ described in equation~\eqref{eq:twisted_dnc}.}
	\label{fig:s10}
\end{figure}

As we have seen, pure ambipolar diffusion leads the magnetic field inside the star to an equilibrium configuration that is a solution of the GS equation. It is interesting to see what happens as we use equilibrium configurations directly derived from the GS equation \eqref{eq:GS} as initial conditions. Hence, we take as our initial condition the dipolar ``twisted-torus'' configuration obtained numerically by \citet{Armaza2015} for
\begin{equation}
	\beta(\alpha)=
	\begin{cases}
		s(\alpha-\alpha_\text{surf})^{1.1} & \text{if}\quad \alpha \geq \alpha_\text{surf} \\
		0 & \text{if}\quad \alpha < \alpha_\text{surf}
	\end{cases} \,,\label{eq:twisted}
\end{equation}
where $\alpha_\text{surf}\equiv\alpha(R,\pi/2)>0$, and $s$ is an amplitude.
The code described in \citet{Armaza2015} uses this $\beta(\alpha)$ relation to solve for $\alpha$ in a self-consistent way from the GS equation, assuming a constant value of $\chi'$.
This equilibrium configuration is displayed in Fig.~\ref{fig:s10}(a).
As can be seen from equation~\eqref{eq:twisted}, in this configuration the toroidal field is explicitly confined to the region where the poloidal field lines close inside of the star. Fig.~\ref{fig:s10}(b) shows the value of $\chi$ which was chosen as a linear function of $\alpha$ so that
\begin{equation}
\chi = b^2\alpha+\chi_0 \,, \label{eq:twisted_dnc}
\end{equation}
where $\chi_0$ is a constant chosen so $\chi$ integrated over the volume of the star is null, as this corresponds to the conservation of the total number of charged particles. 

\begin{figure}
	\centering
	\includegraphics[width=\linewidth]{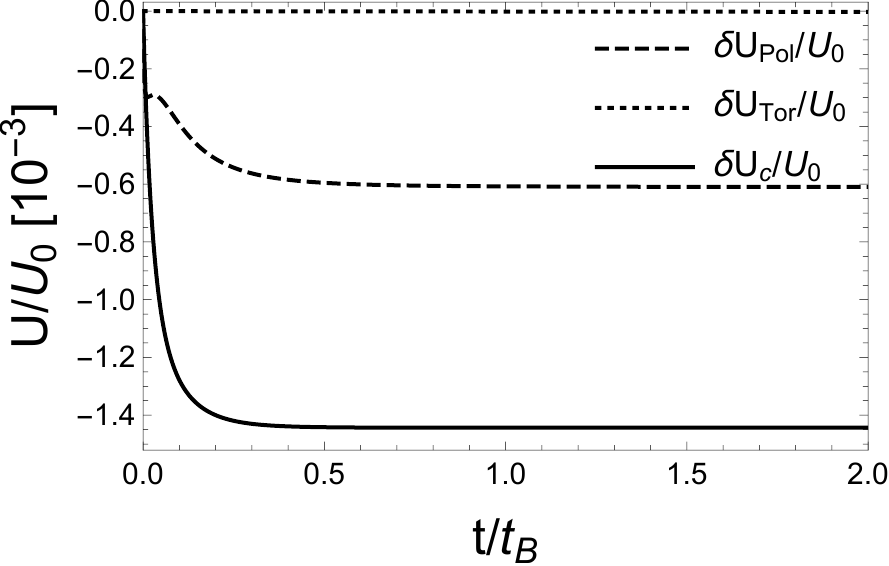}
	\caption{Evolution in time of the variation of the poloidal, toroidal, and particle energy with respect to their initial values for the initial configuration shown in Fig.~\ref{fig:s10}. The simulation was performed for $b^2=0.31$ using a grid of $N_r,N_\theta,N_{\text{Exp}}=100,101,30$, respectively.}
	\label{fig:s10_U}
\end{figure}
\begin{figure}
	\centering
	\includegraphics[width=\linewidth]{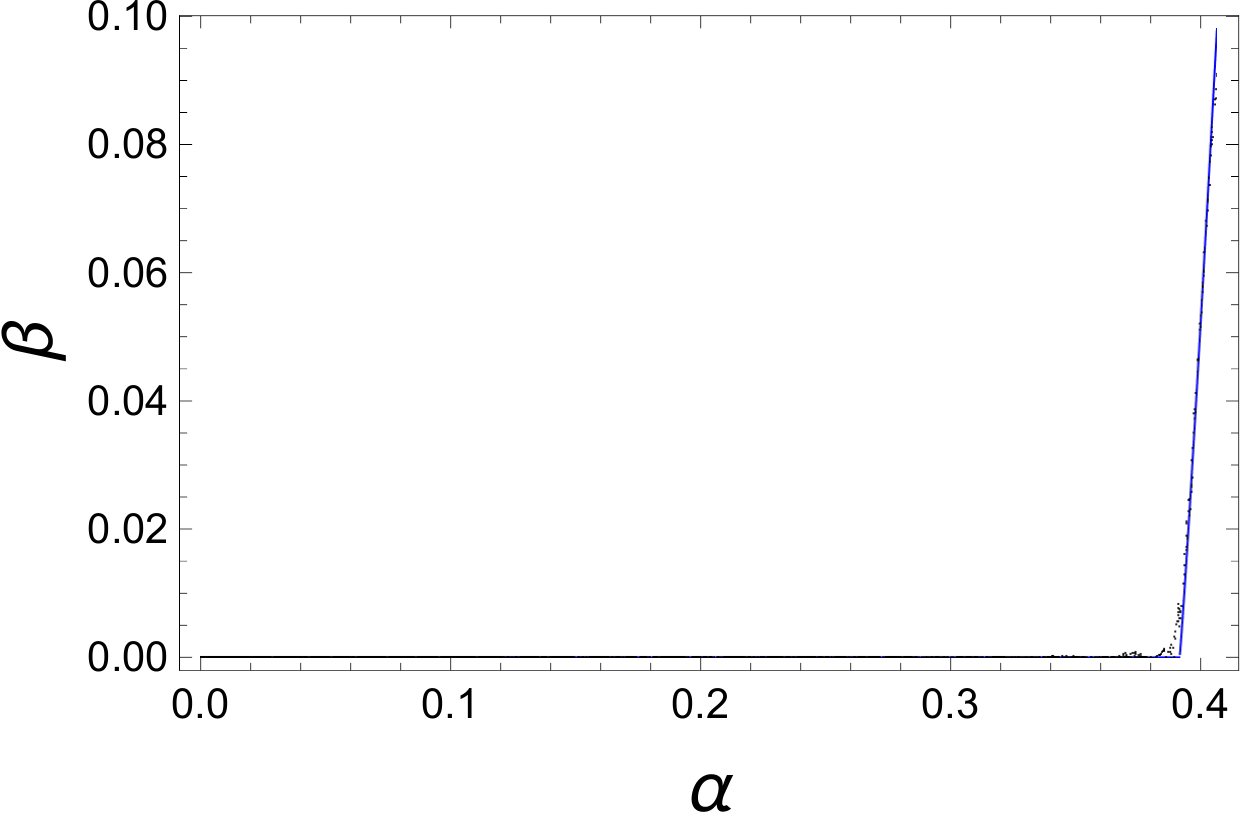}
	\caption{Black dots represent $\beta$ vs. $\alpha$ after $t/t_B=1$ using of the magnetic configuration shown in Fig.~\ref{fig:s10} as initial condition. The solid blue line shows the imposed analytic $\beta(\alpha)$ relation of the magnetic field (equation~\ref{eq:twisted}).}
	\label{fig:s10_vs}
\end{figure}

Taking this as the initial state, Fig.~\ref{fig:s10_U} displays the time-evolution of the variation of each energy component respect to its initial value. We see how both magnetic and particle energies quickly converge to a constant value. However, the relative changes are very small ($\sim 10^{-3}$ for $\delta U_c$ and $\delta U_\text{Pol}$, and $\sim10^{-6}$ for $\delta U_\text{Tor}$), suggesting that the configuration of the field has not changed, as any change in the magnetic field configuration due to ambipolar diffusion has dissipation of energy associated, due to binary collisions. In Fig.~\ref{fig:s10_vs}  we show that the relation $\beta(\alpha)$ at $t=t_B$ is almost identical to the initial one, suggesting that the initial configuration is stable in axial symmetry.

\section{Conclusions and discussion}
\label{sec:conclusions}

We have presented results from simulations that evolve simultaneously and consistently the structure of the magnetic field and the density field of the charged particles (taken to be a locally neutral mix of protons and electrons) inside the core of an isolated NS through ambipolar diffusion in axial symmetry. Following the discussions of \citet{Goldreich1992a}, we modeled the neutrons as a motionless, uniform background that produces a frictional drag on the charged particles and neglected the currents in the crust (assumed to have a very low conductivity), as well as the effects of superfluidity, superconductivity, and weak interactions (``Urca reactions''). We considered different initial conditions, including purely poloidal, purely toroidal, and mixed magnetic field configurations.

The evolution of the magnetic field was found to proceed in two stages, whose time-scales depend on the balance between the frictional drag and other forces. In a short time $\sim 0.1 t_c$ (equation \ref{eq:tc}), the divergence of the ambipolar diffusion velocity perturbs the density of the charged particles, creating pressure gradients that cancel the irrotational part of the magnetic forces. Afterwards, the velocity field is purely solenoidal, but drives a substantial evolution of the magnetic field on a longer time-scale $\sim 0.1 t_B\propto 1/B^2$ (equation \ref{eq:tb}), leading to an equilibrium state that no longer evolves.

In this final equilibrium state, there is a complete balance between the forces due to the magnetic field and the degeneracy pressure gradients, which strongly constrains the magnetic field configuration. The toroidal field is confined to regions of closed poloidal field lines (``twisted tori''), whose number is conserved by the evolution, since ambipolar diffusion by itself cannot produce magnetic reconnections. Outside these tori, i.e. along poloidal field lines that extend outside the star, the toroidal field disappears through ``unwinding'' of the field lines by a toroidal component of the ambipolar velocity. Moreover, these configurations satisfy the Grad-Shafranov (GS) equation \eqref{eq:GS}, because they are hydromagnetic equilibria in a barotropic fluid, namely the homogeneous mixture of protons and electrons. Previously found solutions of the GS equation \citep[e.g.,][]{Armaza2015} are confirmed to be stable in our simulations. 

Several caveats need to be noted and addressed in future work:
\begin{enumerate}
	\item Previous work indicates that there are no stable hydromagnetic equilibria in barotropic stars \citep{Braithwaite2009,Akgun2013,Mitchell2015}, but their main instabilities are non-axisymmetric. Thus, since in the present models the magnetic forces are balanced by the pressure gradient of a barotropic fluid (namely the charged particles), the resulting equilibria are likely to become unstable if considered in full 3 dimensions, which might lead to a complete dissipation of the magnetic field.\\

	\item In real neutron stars, there is a strong radial density gradient, and the neutrons are also allowed to move. Since the density gradients of neutrons and charged particles are different, their velocity fields must also be different, as long as Urca reactions cannot adjust the composition ``in real time'' (see below). Thus, there will always be a relative motion of neutrons and charged particles, with the associated drag force, setting time-scales that might be somewhat shorter, but not very different from the ones found here.\\

	\item Urca reactions (conversions of neutrons into charged particles, and vice-versa, through weak interactions) are highly temperature-dependent and can have two effects. In young, hot NSs, they can adjust the composition of a fluid element while it is pushed by the magnetic forces, therefore all particles can be considered as one fluid, which moves together with the magnetic field on a time-scale set by the weak interactions, avoiding frictional drag. In old, cool NSs (to which the present simulations are more directly applicable), they are too slow to affect the evolutionary time-scales discussed here, but they might set the (extremely long) time-scale for the evolution of the final equilibria found here.\\

	\item If the crust is highly conducting, making its currents evolve more slowly than those in the core, the magnetic field evolution and equilibria in the latter will be more constrained, and the crust will likely set the time-scale for the evolution of the external field.\\

	\item If the charged particles include not just protons and electrons, but also muons or other particle species, with a different density profile, they will no longer act as a barotropic fluid, making the evolution more complicated, probably slower, and perhaps leading to different, less constrained equilibria.\\

	\item If the neutrons in the NS core are superfluid, and/or the protons are superconducting, this should at the very least change their couplings, modifying the time-scales for the processes simulated here \citep[e.g.,][]{Glampedakis2011}. It is not clear yet whether entrainment between neutrons and protons will change the processes in a more fundamental, qualitative way.
\end{enumerate}

\section*{Acknowledgements}

This work was supported by CONICYT doctoral fellowship 21120953 (F.C.),
FONDECYT projects 1150411 (A.R.) and 1150718 (J.A.V.), CONICYT PIA
project ACT1405 and the Center for Astrophysics and Associated
Technologies (CATA; PFB-06). We thank C. Armaza for having provided numerical solutions of the GS equation, which were used as initial conditions for some of our simulations.




\bibliographystyle{mnras}
\bibliography{library} 



\appendix

\section{Numerical code}
\label{app:code}

\subsection{Discretization}

\begin{figure}
\centering
\includegraphics[width=\linewidth]{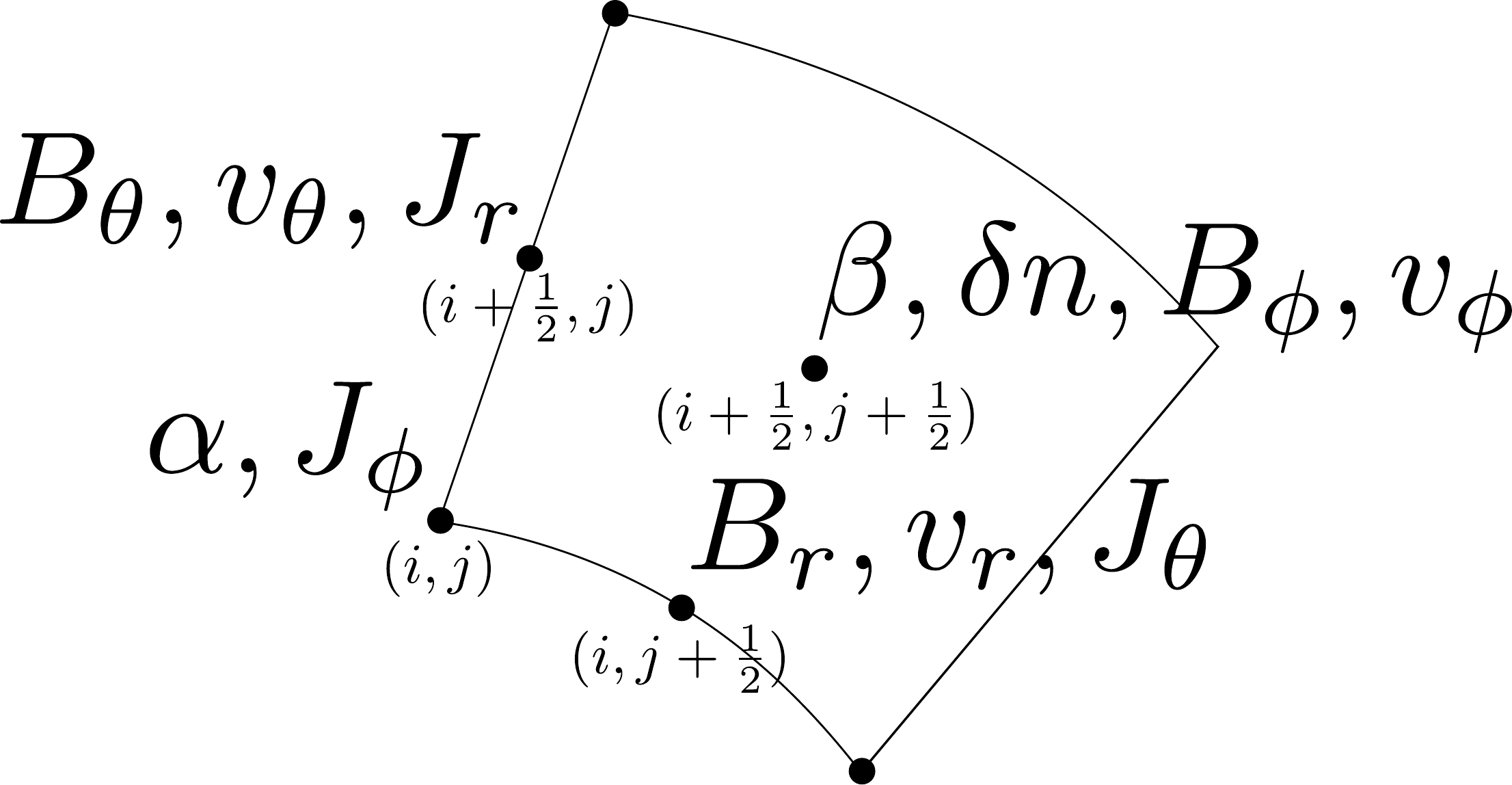}
\caption{Schematic representation of the positions where the given variables are evaluated in the box located at the point $(i,j)$.}
\label{fig:box}
\end{figure}

The grid is composed of $N_r$ points in the radial direction and $N_\theta$ in the polar direction, that are placed in spherical coordinates at
\begin{gather}
r_{i} =\sqrt{\frac{i-1}{N_r-1}} \,,\\
\theta_{j} = (j-1)\Delta\theta\,,
\end{gather}
where $i=1,..,N_r; j=1,..,N_\theta$ and $\Delta \theta = \pi/(N_\theta-1)$. The point located at $(i,j)$ defines a box of size $r_i \leq r \leq r_{i+1}$, $\theta_j \leq \theta \leq \theta_{j+1}$ in which the discretized variables are defined at the positions shown in Fig.~\ref{fig:box}. Here, the variables with index $i+1/2$ are defined at the midpoints $r_{i+1/2} = (r_i+r_{i+1})/2$. The quantities $f(r)$ which are given functions of the radius, known at all points of the grid; such as  $n_n$
, $n_c$, $\gamma_{ij}$, and $K_{cc}$; are discretized as $f_{i}=f(r_i)$.

\subsection{Numerical divergence}

As the code aims to conserve the number of charged particles and the
toroidal flux, the definition of the numerical divergence is
relevant. Hence, the fluxes
over each face of the box defined at $(i,j)$ are taken as the local value of the
normal component of the vector field $\Vect{F}$ times the surface area of
that face. As a result, the value of the divergence at the center of
the box $ (\Div \Vect F)_{i,j} =\left.\Div \Vect
F\right|_{r_{i+1/2},\theta_{j+1/2}}$ is given by
\begin{equation}
\begin{split}
(\Div \Vect F)_{i,j} =&\, 3\frac{r_{i+1}^2F_r^{i+1,j}-r_{i}^2F_r^{i,j}}{r_{i+1}^3-r_{i}^3} \\
&+\frac{3(r_{i+1}^2-r_{i}^2)}{2(r_{i+1}^3-r_{i}^3)}\frac{\sin\theta_{j+1}F_\theta^{i,j+1}-\sin\theta_{j}F_\theta^{i,j}}{\cos\theta_j-\cos\theta_{j+1}} \label{eq:divnumerica}\,.
\end{split}
\end{equation}

\subsection{Evaluation of $B$ and $J$}

The discretization of $\Vect{B}$ that is consistent with the condition $(\Div \Vect B)_{i,j}=0$, when using the
definition taken for the numerical divergence, is given by
\begin{align}
B_r^{i,j} &= \left(\frac{1}{r^2 \sin\theta}\frac{\partial\alpha}{\partial\theta}\right)_{i,j+1/2}
= \frac{1}{r_i^2}\frac{\alpha_{i,j+1}-\alpha_{i,j}}{\cos\theta_j-\cos\theta_{j+1}} \,,\\
B_\theta^{i,j} &= -\left(\frac{1}{r \sin\theta}\frac{\partial\alpha}{\partial r}\right)_{i+1/2,j}
=-\frac{2}{\sin\theta_j}\frac{\alpha_{i+1,j}-\alpha_{i,j}}{r_{i+1}^2-r_{i}^2} \,,\\
B_\phi^{i,j} &= \frac{\beta_{i,j}}{r_{i+1/2} \sin\theta_{j+1/2}} \,.
\end{align}

We take a similar discretization for the electric current density since it also has zero divergence, namely,		
\begin{align}
J_r^{i,j}      &=\left(\frac{1}{r^2 \sin\theta}\frac{\partial\beta}{\partial\theta}\right)_{i+1/2,j}=\frac{1}{r_{i+1/2}^2}\frac{\beta_{i,j}-\beta_{i,j-1}}{\cos\theta_{j-1/2}-\cos\theta_{j+1/2}} \,,\\
J_\theta^{i,j} &= -\left(\frac{1}{r \sin\theta}\frac{\partial\beta}{\partial r}\right)_{i,j+1/2} =-\frac{2}{\sin\theta_{j+1/2}}\frac{\beta_{i,j}-\beta_{i-1,j}}{r_{i+1/2}^2-r_{i-1/2}^2} \,.
\end{align}

The discretization used for both $J_r$ and $J_\theta$ is equivalent
to evaluating the current using Stokes' theorem, so that the integrals over the loops are done taking the value of the field as constant along each segment of
the loop. The same is done to compute the $J_\phi$
component, namely,
\begin{equation}
\begin{split}
J_\phi^{i,j} =& -\frac{1}{\frac{1}{2}(r_{i+1/2}^2-r_{i-1/2}^2)\Delta\theta}\Big[B_r^{i,j}(r_{i+1/2}-r_{i-1/2}) - B_\theta^{i,j}r_{i+1/2}\Delta\theta\\
&- B_r^{i,j-1}(r_{i+1/2}-r_{i-1/2}) + B_\theta^{i-1,j}r_{i-1/2}\Delta\theta \Big] \,.
\end{split}
\end{equation}

The evaluation of the current at the surface involves the evaluation
of a phantom point $\alpha_{N_r+1,j}$, which is computed using the
multipolar expansion outside.

\subsection{External magnetic field}

Evaluating the external field involves calculating the multipolar expansion coefficients given by equation~\eqref{eq:a_l}, which is discretized as
\begin{align}
a_l &=-\frac{2l+1}{2(l+1)}\sum_{j=1}^{N_\theta-1} B^{N_r,j}_r \int_{\theta_j}^{\theta_{j+1}} P_l(\cos\theta) \sin\theta  \,d\theta   \,,\\
&=-\frac{2l+1}{2(l+1)}\sum_{j=1}^{N_\theta-1} B^{N_r,j}_r \left(-\frac{P_{l+1}(\cos\theta)-P_{l-1}(\cos\theta)}{2l+1}\right)_{\theta_j}^{\theta_{j+1}}  \,,\\
&=\frac{1}{2(l+1)}\sum_{j=1}^{N_\theta-1} B^{N_r,j}_r \left(P_{l+1}(\cos\theta)-P_{l-1}(\cos\theta)\right)_{\theta_j}^{\theta_{j+1}}	  \,,
\end{align}
for $l=2,3,...,N_{\text{Exp}}$, where $N_{\text{Exp}}$ is an integer number giving the number of multipoles used.

We need a phantom point $\alpha_{N_r+1,j}$ to evaluate $J_\phi$ at the surface, but this value is fixed by the condition that $B_\theta$ must also be continuous at the surface, which leads to 
\begin{align}
B_\theta(1,\theta_j) &= -\frac{2}{\sin\theta_j}\frac{\alpha_{N_r+1,j}-\alpha_{N_r-1,j}}{r_{N_r+1}^2-r_{N_r-1}^2} 
\\&= \sum_{l=1}^{N_{\text{Exp}}} a_l P_l^1(\cos\theta_j)  \,.
\end{align}
Hence, the phantom point is taken as
\begin{equation}
\alpha_{N_r+1,j} = \alpha_{N_r-1,j}-	\left(r_{N_r+1}^2-r_{N_r-1}^2\right)\frac{\sin\theta_j}{2}\sum_{l=1}^{N_{\text{Exp}}} a_l P_l^1(\cos\theta_j)  \,.
\end{equation}

\subsection{Ambipolar velocity}

The velocities are calculated as 
\begin{equation}
v_{A,r}^{(i,j)} = b^2\left(J_\theta^{(i,j)}\bar B_\phi - \bar J_\phi \bar B_\theta \right)  - \left(\frac{\partial\chi}{\partial r}\right)_{i,j+1/2}  \,,
\end{equation}
where the numerical density perturbations are defined at $\chi_{i,j} =\chi(r_{i+1/2},\theta_{j+1/2})$, so that
\begin{equation}
\left(\frac{\partial\chi}{\partial r}\right)_{i,j+1/2} =  \frac{\chi_{i,j}-\chi_{i-1,j}}{r_{i+1/2}-r_{i-1/2}}  \,.
\end{equation}	
Hereafter, the ``bar'' values; such as  $\bar B_\theta$, $\bar
B_\phi$, and $\bar J_\phi$; correspond to linear
interpolation of the different variables taking the closest neighbors
to the point in which the quantity, in this case $v_{A,r}$, is being
evaluated.

At the surface of the star, we are assuming that $v_{A,r}=0$. For the components $v_\theta$ and $v_\phi$ we have
\begin{gather}
v_{A,\theta}^{(i,j)} = b^2\left(\bar J_\phi \bar B_r - J_r^{(i,j)}\bar B_\phi \right) -\frac{\chi_{i,j}-\chi_{i,j-1}}{r_{i+\frac12}\Delta\theta} \,,\\
v_{A,\phi}^{(i,j)} =b^2\left(\bar J_r \bar B_\theta - \bar J_\theta \bar B_r \right).
\end{gather}

\subsection{Time evolution}

Knowing the numerical value of the ambipolar velocity, the evaluation of the time derivative of $\alpha$ is
\begin{equation}
\left( \frac{\partial\alpha}{\partial t}\right)_{i,j}=r_i\sin\theta_j\left( \bar v_{A,r} \bar B_\theta - \bar v_{A,\theta} \bar B_r \right)  \,,
\end{equation}
from which we can evolve $\alpha$ using second-order Runge-Kutta to evolve from the instant $t_s$ to $t_{s+1}\equiv t_s+\Delta t$:
\begin{gather}
\alpha_{i,j}\left(t_s+\frac{\Delta t}{2}\right) = \alpha_{i,j}(t_s) + \frac{\Delta t}{2}\left( \frac{\partial\alpha}{\partial t}\right)_{i,j}(t_s) \,,\\
\alpha_{i,j}(t_s+\Delta t) = \alpha_{i,j}(t_s) + \Delta t\left( \frac{\partial\alpha}{\partial t}\right)_{i,j}\left(t_s+\frac{\Delta t}{2}\right) \,.
\end{gather}
This numerical method is used to evolve all quantities in time.

To evolve the toroidal field, we first define the following quantities 
\begin{gather}
F_r = r\left(B_r v_{A,\phi} -B_\phi v_{A,r} \right) \label{Fr}  \,,\\
F_\theta = B_\phi v_{A,\theta} -B_\theta v_{A,\phi} \label{Fth}  \,.
\end{gather}
These quantities are proportional to the fluxes in
equation~\eqref{eq:evolucionbeta}, and are discretized as
$F_r^{i,j}=F_r(r_{i},\theta_{j+1/2})$ and
$F_\theta^{i,j}=F_\theta(r_{i+1/2},\theta_{j})$, where
\begin{gather}
F_r^{(i,j)} = r_i\left(B_r^{(i,j)} \bar v_{A,\phi} - \bar B_\phi \bar v_{A,r} \right)  \,,\\
F_\theta^{(i,j)} = \bar B_\phi \bar v_{A,\theta} - B_\theta^{(i,j)}\bar v_{A,\phi}   \,.
\end{gather}

Then, using the definition of numerical divergence from equation~\eqref{eq:divnumerica} along with equation~\eqref{eq:evolucionbeta}, it is possible to evaluate the time derivative of $\beta$ in a conservative manner as
\begin{equation}
\left( \frac{\partial\beta}{\partial t} \right)_{i,j} =f_{i,j} (F_r^{(i+1,j)}- F_r^{(i,j)}) - g_{i,j} (F_\theta^{(i,j+1)} - F_\theta^{(i,j)}) \label{eq:disc_Beta}\,,
\end{equation}
where the geometrical factors $f_{i,j}$ and $g_{i,j}$ are
\begin{gather}
f_{i,j} = \sin\theta_{j+1/2}\frac{3 r^2_{i+1/2}}{r_{i+1}^3-r_{i}^3} \,,\\
g_{i,j} = \frac{3 r_{i+1/2}(r_{i+1}^2-r_{i}^2)}{2(r_{i+1}^3-r_{i}^3)} \frac{\sin^2\theta_{j+1/2}}{\cos\theta_{j}-\cos\theta_{j+1}} \,.
\end{gather}

Having the numerical value of $\Vect{v_A}$, the evolution of the
perturbation to the density of charged particles is direct.
Using equation~\eqref{eq:continuidad_a}, we have	
\begin{equation}
\left(\frac{\partial\chi}{\partial t}\right)_{i,j} =  -\left(\Div\Vect{v_A}\right)_{i,j} \,,
\end{equation}
where the numerical divergence defined in equation~\eqref{eq:divnumerica} has to be used. 

\section{Numerical Tests}
\label{app:tests}

\subsection{Purely poloidal field}
\begin{figure}
\centering
\includegraphics[width=\linewidth]{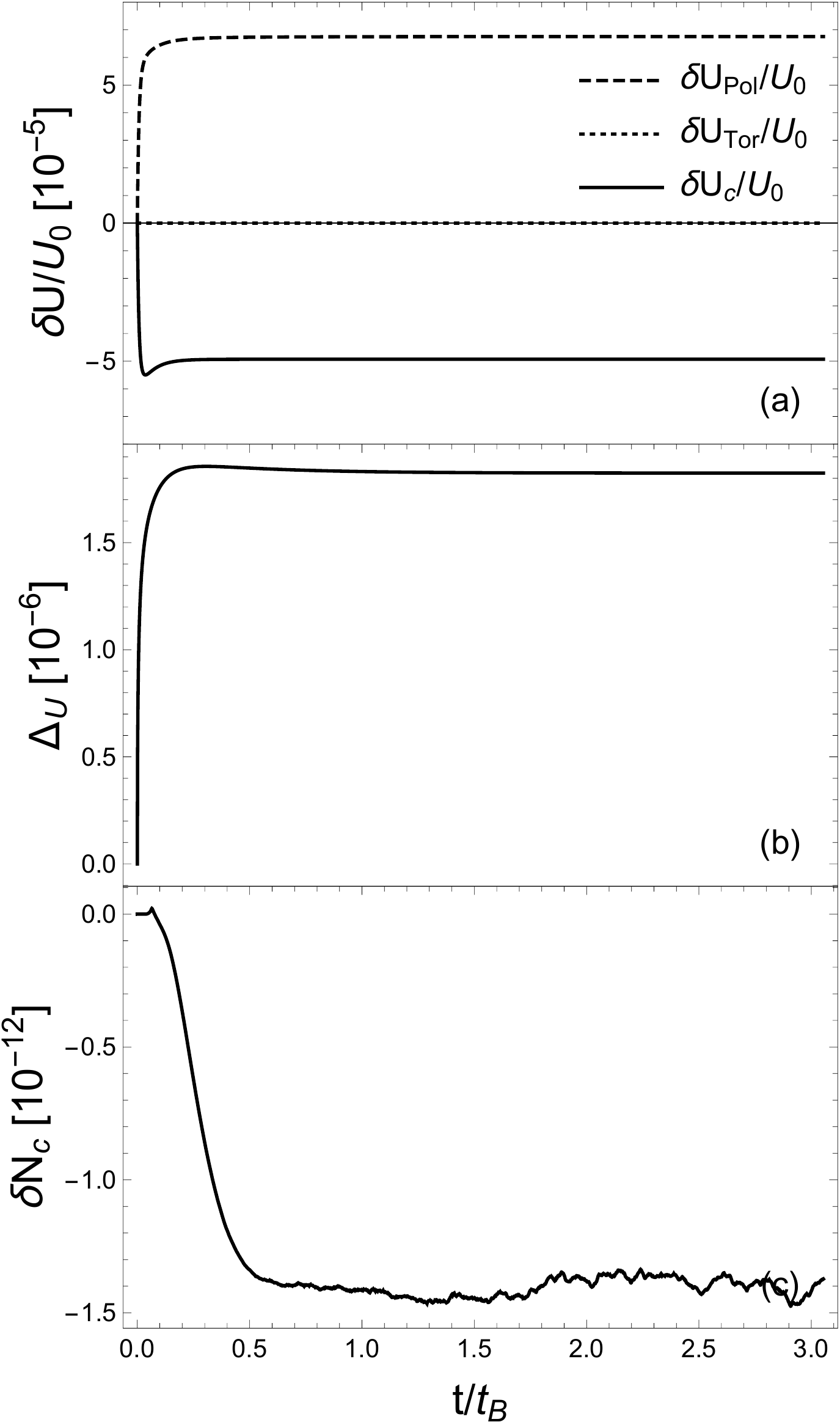}
\caption{Simulation of a purely poloidal field defined by equation~\eqref{eq:b1} and equation~\eqref{eq:b2}, in which we have taken $b^2=0.14$ and a resolution of $N_r=60, N_\theta=91$, with $N_{\text{Exp}}=27$ multipoles. In the figures we see: (a) Evolution of the relevant energies in time. (b) Conservation of total energy. (c) Variation of the charged particle number.}
\label{fig:eq.pol}
\end{figure}

In order to test the reliability of the code, we compared it with
some analytical solutions of the GS equation~\eqref{eq:GS} adapted to our ambipolar diffusion setting. As shown by \citet{Gourgouliatos2013} an analytic solution to this equation for a uniform density profile is 
\begin{gather}
\alpha(r,\theta) = \alpha_0\left( 5r^2 -3r^4\right) \sin^2\theta \,, \label{eq:b1}\\
\beta(r,\theta) = 0 \,,\\
\chi(r,\theta) = 30 b^2 \alpha_0\alpha(r,\theta)\,, \label{eq:b2}
\end{gather}
where $\alpha_0$ is an amplitude. It can be seen that a linear
relation between $\alpha$ and $\chi$ has been imposed in this
solution. This configuration has been used as an input for one of our
simulations. The system is evolved during $3t_B$. The evolution of the variation of the different energies involved, respect to their initial values is shown in Fig.~\ref{fig:eq.pol}(a).
Here it can be seen that only very small changes in the energies are present, as expected from an initial equilibrium configuration. These small variations are expected due
to the numerical precision of the discretization, also the change of the energies respect to their initial values is of the order of $10^{-5}$ which is negligible.

An important test of any numerical code is to see if it can properly conserve the total energy. The total energy loss due to collisions $U_{\text{Col}}$
can be computed integrating equation \eqref{eq:dUcol} in time, as the
simulation runs. Then the quantity $U=U_B+U_c+U_{\text{Col}}$ must be
conserved in the simulations. Fig.~\ref{fig:eq.pol}(b) shows the
relative change of $U$, namely, 
\begin{equation}
\Delta_U=\left|U/U(0)-1\right| \,.
\end{equation}
It can be seen that the energy is conserved within $\Delta_U\sim10^{-6}$. Similarly, Fig.~\ref{fig:eq.pol}(c) shows the variation of the number of charged particles $\Delta N_c$,
computed integrating $\chi$ over the volume of the star. As
can be seen, the variation of the number of charged particles in time is small, $\sim
10^{-12}$, which is consistent with the fact that the code was built
to be conservative for $\chi$.

In order to check the magnitude of the local numerical errors of our simulations we have defined the quantity
\begin{equation}
\text{Err}(\alpha,t_f)=\frac{\max_{i,j}|\alpha_{i,j}(t=t_f)-\alpha_{i,j}(t=0)|}{\max_{i,j}|\alpha_{i,j}(t=0)|}\,.
\end{equation}
In this simulation we see local errors of magnitude $\text{Err}(\alpha,3t_B)=1.3\times10^{-4}$ and $\text{Err}(\chi,3t_B)=2.6\times10^{-3}$, which are small.

\subsection{Mixed field configuration}

\begin{figure}
\centering
\includegraphics[width=\linewidth]{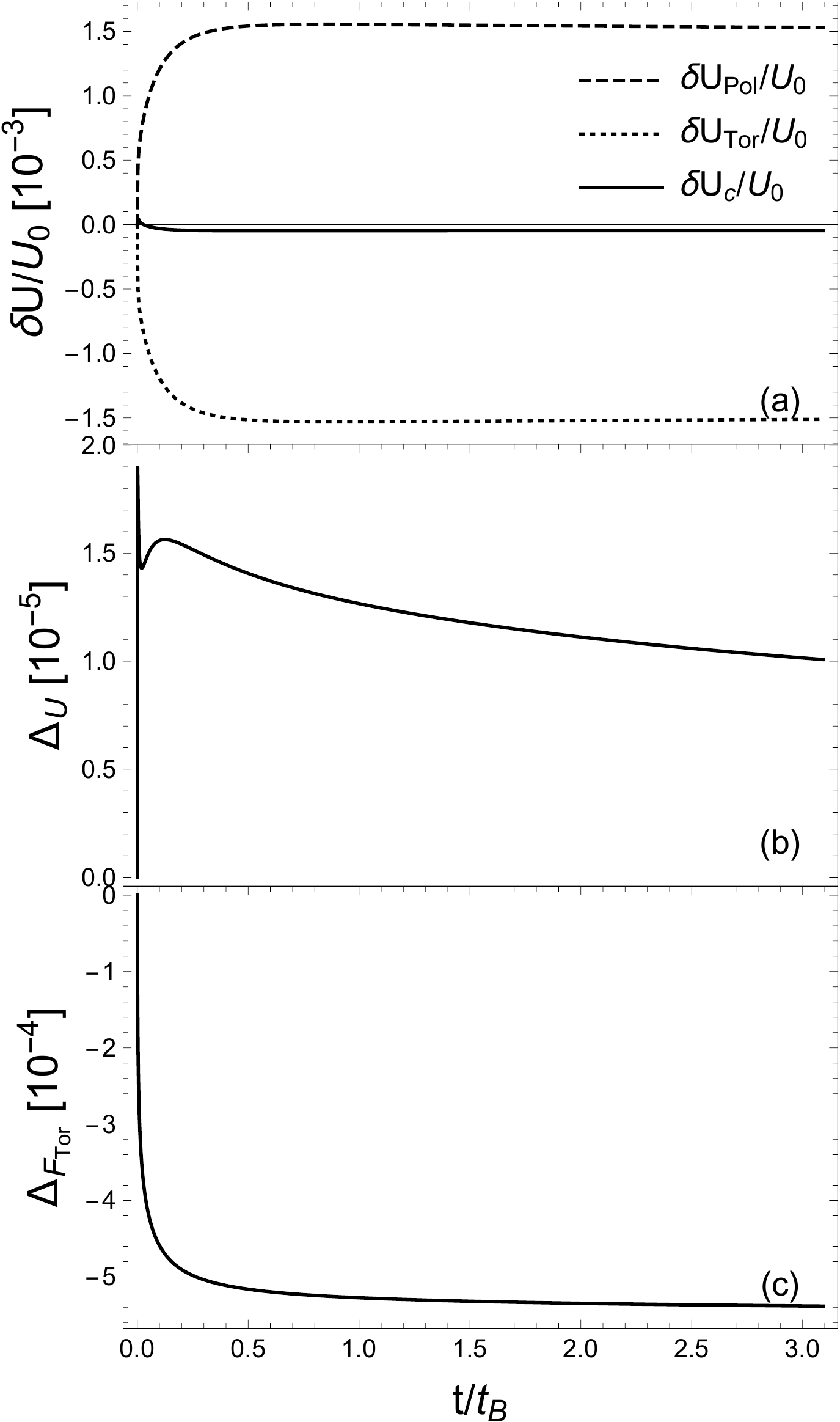}
\caption{Simulation of a mixed field defined by equation~\eqref{eq:b8}, \eqref{eq:b9}, and \eqref{eq:b10}, in which we have taken $b^2=0.14$ and a resolution of $N_r=60, N_\theta=91$, with $N_{\text{Exp}}=27$ multipoles. In the figures we see (a) evolution of the relevant energies in time. (b) Conservation of total energy. (c) Conservation of toroidal flux.}
\label{fig:eq.pol.tor}
\end{figure}

Another analytic solution of the GS equation found by \citet{Gourgouliatos2013}, adapted to our ambipolar diffusion setting, is given by
\begin{gather}
\alpha = \frac{\alpha_0}{s^2}\left( r\frac{j_1(sr)}{j_1(s)}-r^2\right) \sin^2\theta \, \label{eq:b8}\,,\\
\beta = s\alpha \,\label{eq:b9}\,,\\
\chi = \alpha_0b^2\alpha \,,\label{eq:b10}
\end{gather}
where $s=5.763$, and $j_1(r)$ is a spherical Bessel function with
index 1. This case corresponds to a mixed poloidal and toroidal magnetic field confined in the star.
This configuration has been used as an input for one of our simulations.
The system is evolved during $3t_B$.
In this case, as shown in Fig.~\ref{fig:eq.pol.tor}(a), there are some variations in the energies at the beginning of the simulations, mostly toroidal energy
being converted to poloidal energy, but the variations are small, and,
as shown in Fig.~\ref{fig:eq.pol.tor}(b), these variations
conserve the energy to a good approximation.
Fig.~\ref{fig:eq.pol.tor}(c) shows the relative variation of the toroidal magnetic flux $F_{\text{Tor}}$ in time, defined as $\Delta_{F_{\text{Tor}}}=F_{\text{Tor}}/F_{\text{Tor}}(0)-1$.
This variable is conserved in the case of purely toroidal
fields, however, in our case the flux decreases at the beginning of
the simulation, which is in agreement with the information displayed
in Fig.~\ref{fig:eq.pol.tor}(a) In our simulation, the
variation of the number of charged particles is never
larger than $2\times10^{-12}$.

\begin{figure}
\centering
\includegraphics[width=\linewidth]{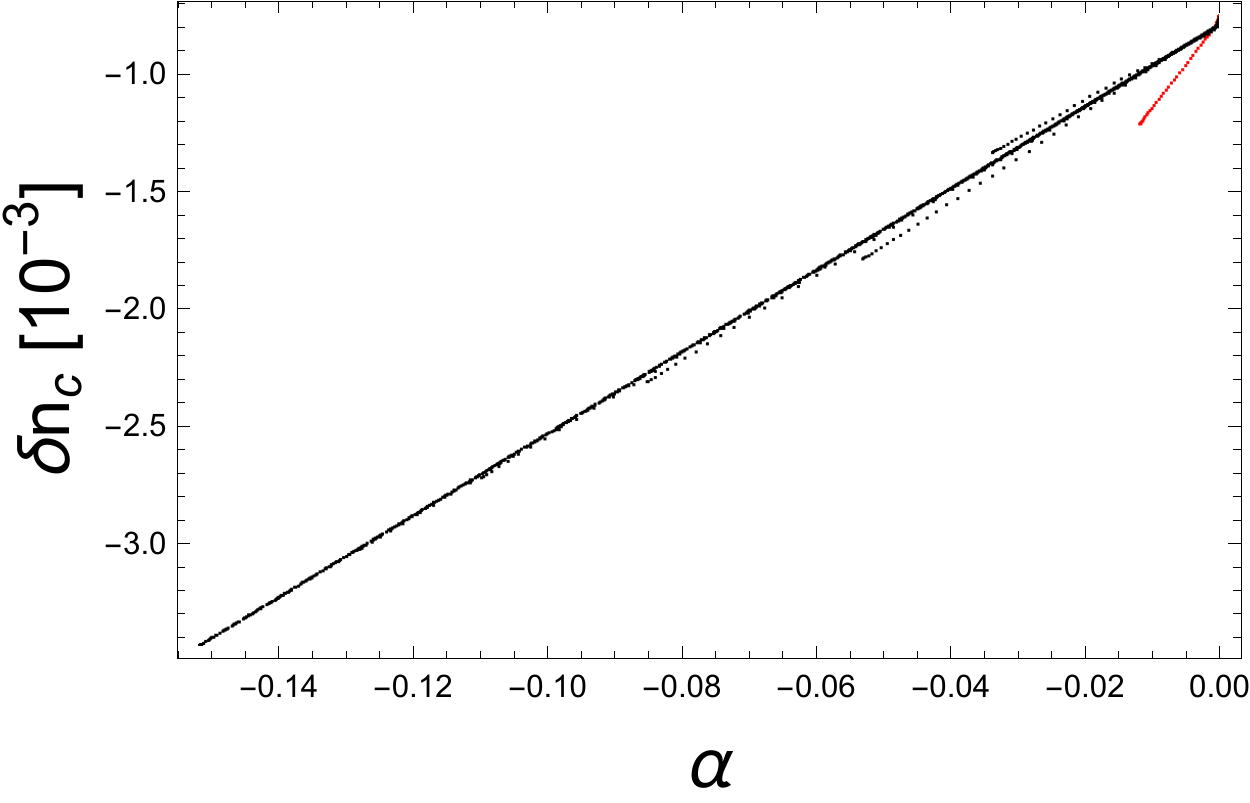}
\caption{$\chi$ vs. $\alpha$ after $3t_B$ for the simulation described in Fig.~\ref{fig:eq.pol.tor}.}
\label{fig:eq.pol.tor.dnc}
\end{figure}

Fig.~\ref{fig:eq.pol.tor.dnc} shows the $\chi(\alpha)$
relation after $3t_B$. As can be seen, it keeps the same slope that
it had at $t=0$, meaning that the configuration of the field
has not changed significantly, however there are some points in the figure with a slope different from the initial one. This is a product of the
discretization. To exemplify this, in the figure, the points within
$0<r<0.1$ have been painted red.
These points, at the top right of the figure, have a slope
much bigger than the one they had initially, because they correspond
to the region of the grid in which the radial resolution is the
lowest. To speed up the integration time, we have chosen a
slightly non-uniform grid that has a larger radial separation
close to the center of the star to prevent the Courant time-step
to be prohibitively low. Therefore, we expect a relatively bigger
numerical error there due to a larger radial step.

In order to check the magnitude of the local numerical errors of our simulations we look at the value of the maximum local error of the poloidal component $\text{Err}(\alpha,3t_B)=6\times10^{-3}$. For the toroidal component we have $\text{Err}(\beta,3t_B)=7\times10^{-3}$ and for the particle density perturbation we find $\text{Err}(\chi,3t_B)=0.09$. The higher error in $\chi$ can also be explained in terms of the bigger radial resolution close the center of the star.

\subsection{Non-equilibrium configurations}
\label{app:tests-noneq}

\begin{figure}
\centering
\includegraphics[width=\linewidth]{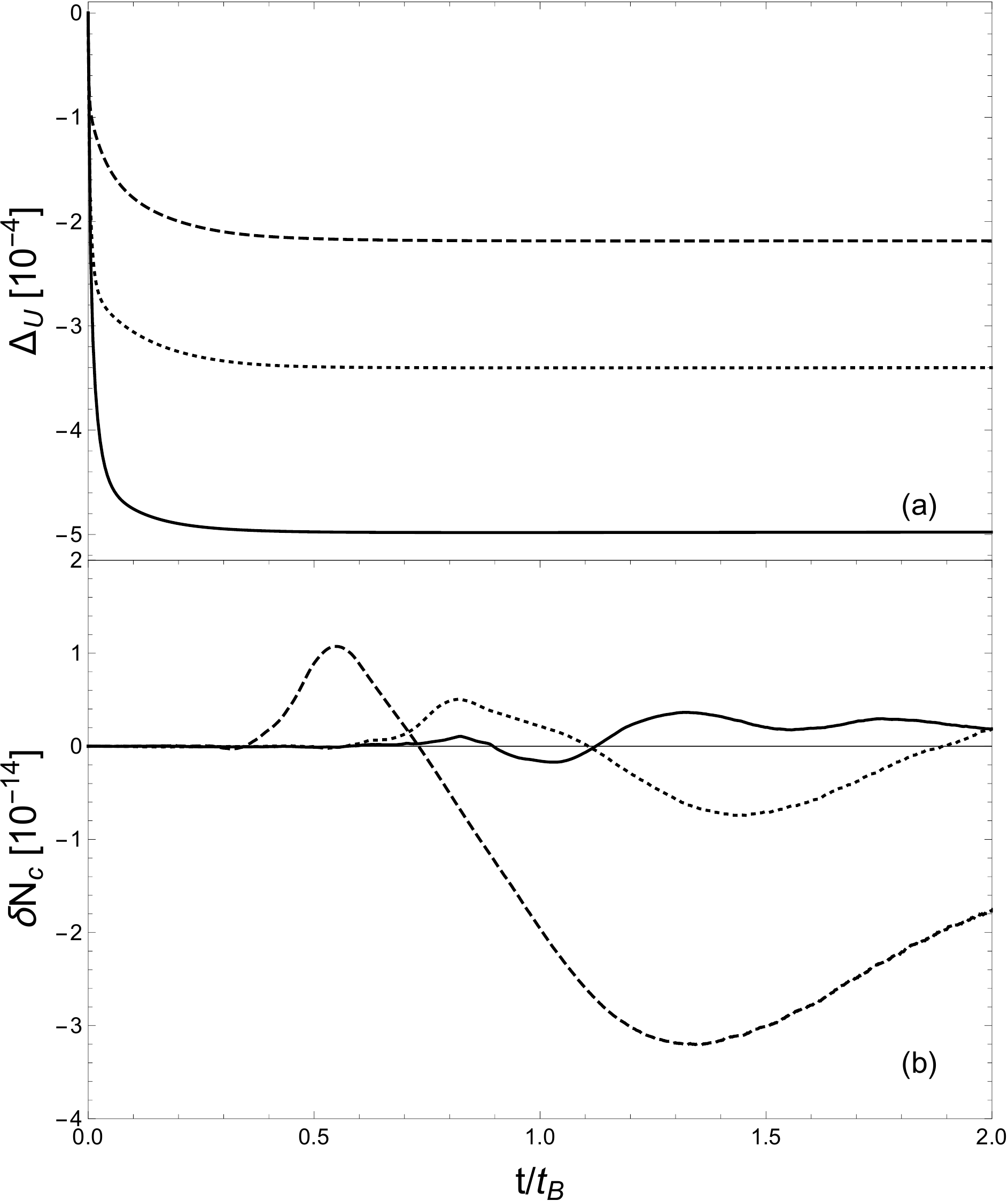}
\caption{(a) Time evolution of the conservation of total energy for the purely poloidal field defined in Fig.~\ref{fig:alfa_puro}, with $b^2=0.03$(dashed), $b^2=0.14$(dotted), $b^2=0.31$(continuous line). (b) Variation of the number of charged particles for the same simulation.}
\label{fig:compb_UN}
\end{figure}

To test the accuracy of the program we also test conservation of quantities in non-equilibrium (and thus evolving) configurations. In the case of the purely poloidal field described in Fig.~\ref{fig:alfa_puro}, we test the conservation of energy and the conservation of the particle number. Fig.~\ref{fig:compb_UN}(a) shows that in the simulation the total energy is conserved within a range of $10^{-4}$ with respect to its initial value. Fig.~\ref{fig:compb_UN}(b) shows how particle number is again conserved within a magnitude of $10^{-14}$.

\begin{figure}
\centering
\includegraphics[width=\linewidth]{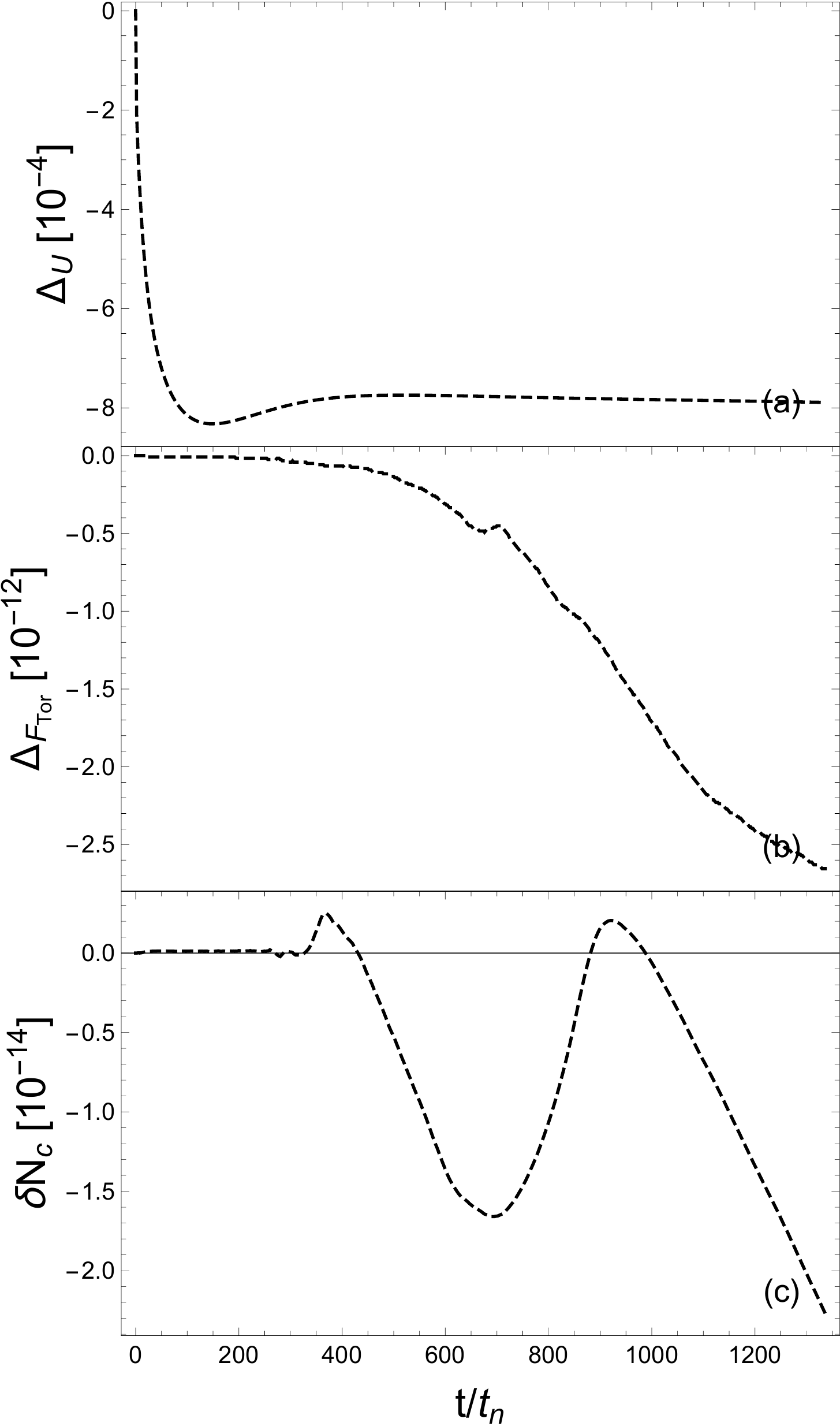}
\caption{(a) Time evolution of the conservation of total energy for the purely toroidal field defined in Fig.~\ref{fig:beta_puro}. (b) Conservation of toroidal flux relative to its initial value. (c) Variation of the number of charged particles for the same simulation.}
\label{fig:beta_vs_UFN}
\end{figure}

\begin{figure}
\centering
\includegraphics[width=\linewidth]{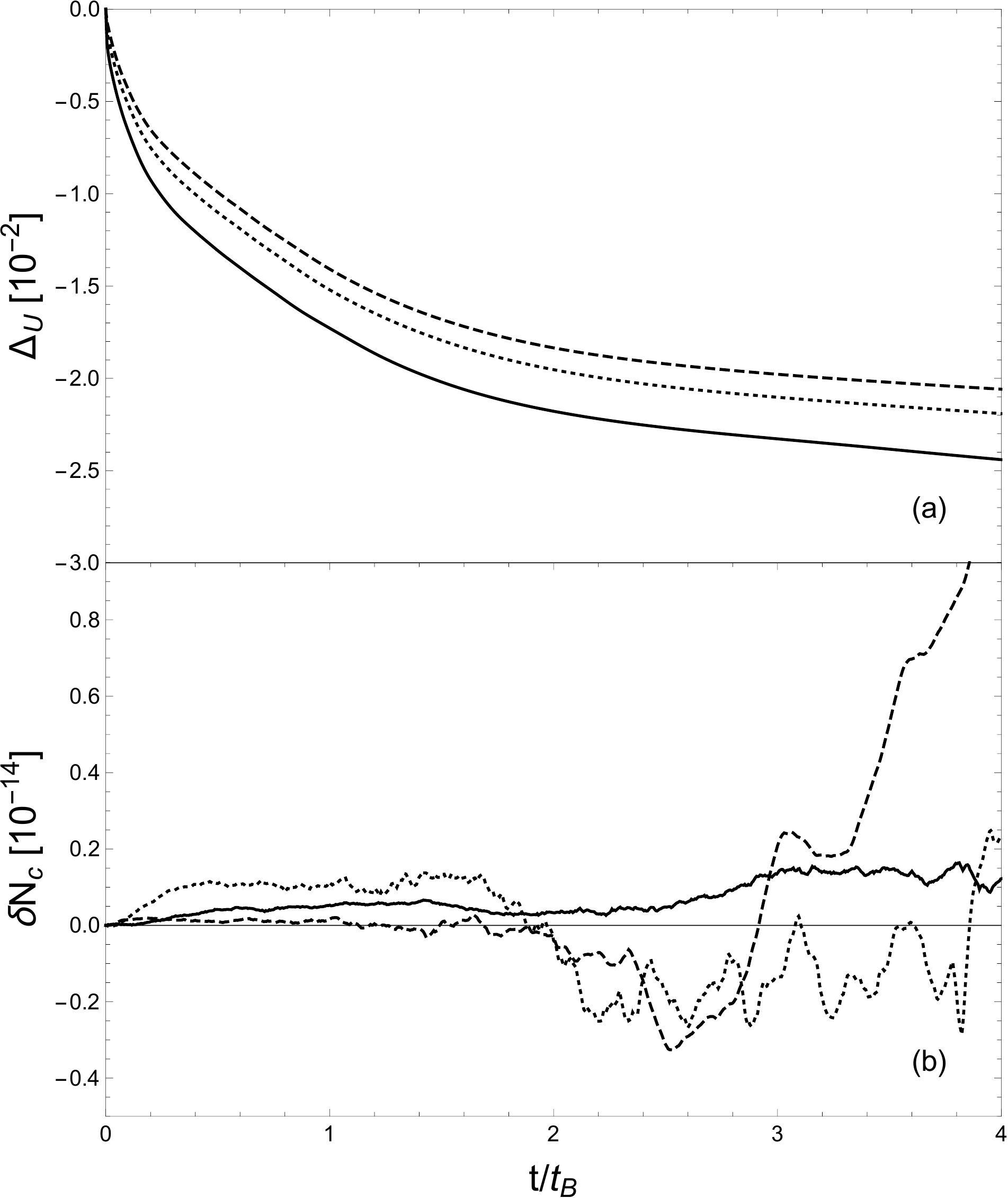}
\caption{(a) Time evolution of the conservation of total energy for the purely poloidal field defined in Fig.~\ref{fig:pol_tor_raro}, with $b^2=0.03$(dashed), $b^2=0.14$(dotted), $b^2=0.31$(continuous line). (b) Variation of the number of charged particles for the same simulation.}
\label{fig:pol_tor_raro_compb_UN}
\end{figure}

The same tests are performed for the purely toroidal field described in Fig.~\ref{fig:beta_puro}. Fig.~\ref{fig:beta_vs_UFN}(a) shows how the total energy is conserved within a range of $10^{-4}$ with respect to its initial value. As can be seen in Fig.~\ref{fig:beta_vs_UFN}(c) particle number is, as expected, conserved within a magnitude of $10^{-14}$. In the case of a purely toroidal field, the toroidal flux is a quantity that must also be conserved, as there is no component $v_{A\phi}$ to twist magnetic field lines to the poloidal direction. Fig~\ref{fig:beta_vs_UFN}(b) shows the evolution of the relative variation of $F_{\text{Tor}}$. As can be seen, the toroidal flux is conserved within a magnitude of $10^{-14}$ relative to its initial value. This is due to the conservative discretization used in equation~\eqref{eq:disc_Beta}. In the case of the mixed poloidal-toroidal field described in Fig.~\ref{fig:pol_tor_raro}, we see in Fig.~\ref{fig:pol_tor_raro_compb_UN} how the total energy is conserved within a range of 1\% and, the particle number is conserved within a magnitude of $10^{-14}$.

\bsp	
\label{lastpage}
\end{document}